\documentclass[submission, Phys]{SciPost}
\usepackage{graphicx}
\usepackage{amsmath,amssymb}
\usepackage{epsfig}
\usepackage{lineno}
\usepackage{braket}
\usepackage{mathtools}

\usepackage{dsfont}
\usepackage[normalem]{ulem}
\usepackage{xcolor}
\usepackage{color}
\usepackage{float}
\usepackage{cancel} 
\usepackage[rightcaption]{sidecap}

\newcommand\be{\begin{equation}}
\newcommand\ee{\end{equation}}
\newcommand{\p}{\partial}
\newcommand{\dr}{^\text{dr}}
\DeclareMathAlphabet{\mathsfit}{T1}{\sfdefault}{\mddefault}{\sldefault}
\SetMathAlphabet{\mathsfit}{bold}{T1}{\sfdefault}{\bfdefault}{\sldefault}

\begin{document}
\graphicspath{{figures/}}
\begin{center}{\Large \textbf{
 Quasiparticle Diffusion for the Toda Fluid in Equilibrium
}}\end{center}

\begin{center}
Seema Chahal\textsuperscript{1*},
Alberto Brollo\textsuperscript{2,3*},
Indranil Mukherjee\textsuperscript{1},
Abhishek Dhar\textsuperscript{1},
Anupam Kundu\textsuperscript{1},
Herbert Spohn\textsuperscript{2,3}
\end{center}

\begin{center}
{\bf 1} International Centre for Theoretical Sciences, Tata Institute of Fundamental Research, Bangalore 560089, India
\\
{\bf 2} Technical University of Munich, CIT, Department of Mathematics, Boltzmannstraße 3, 85748 Garching, Germany
\\
{\bf 3} Technical University of Munich, Department of Physics, James-Franck-Straße 1, 85748 Garching, Germany
\\
* seema.s@icts.res.in, alberto.brollo@tum.de
\end{center}

\begin{center}
\today
\end{center}

\section*{Abstract}

Many-body integrable systems can be understood as a gas of quasiparticles. They propagate ballistically and drive large-scale transport. However, with the exception of the hard rods system, no tools have been available to numerically track such quasiparticles. Focusing on the Toda fluid, whose integrability relies on the availability of a Lax pair, we present a numerical scheme to track quasiparticle trajectories as determined by the time-dependent   eigenvectors of the Lax matrix. Simulating the Toda fluid in thermal equilibrium, this tracking scheme is used to numerical confirm  Brownian motion of a quasiparticle. Simulated is also the motion of a tagged particle. Our numerical results for the diffusion constant matches with a novel TBA prediction. We believe our numerical scheme can be extended to other classical many-particle models possessing a Lax matrix.

\vspace{10pt}
\noindent\rule{\textwidth}{1pt}
\tableofcontents\thispagestyle{fancy}
\noindent\rule{\textwidth}{1pt}
\vspace{5pt}

\section{Introduction}
\label{sec1}
Since Einstein's brilliant discovery of Brownian motion, the dynamics  of an object immersed in 
a fluid is a standard task in Statistical Physics. Einstein investigated the motion of a large object 
suspended in the fluid. But also the dynamics of a tagged  particle of the fluid itself is a well-studied problem. 
Of course, primarily properties of the tracer particle are of interest. But the tracer can also be used to 
probe the random motions of the fluid. There are endless variations. But over the past ten years a new class of fluids, namely \emph{integrable} fluids, have been investigated  in great detail. These are fine-tuned systems which have an extensive number of conservation
laws. Therefore their dynamical properties  are very different from those of a nonintegrable simple fluid. For example, the  Euler equations of an integrable fluid consist of a coupled system of an infinite number of local nonlinear conservation laws. The steady states are generalized Gibbs ensembles (GGE).
Hence local equilibrium  turns into local GGE. In our contribution we study specifically the Toda fluid as a prime example of an integrable fluid. 

\begin{figure}[t]
    \centering
    \includegraphics[width=7.2cm, height=6.2cm]{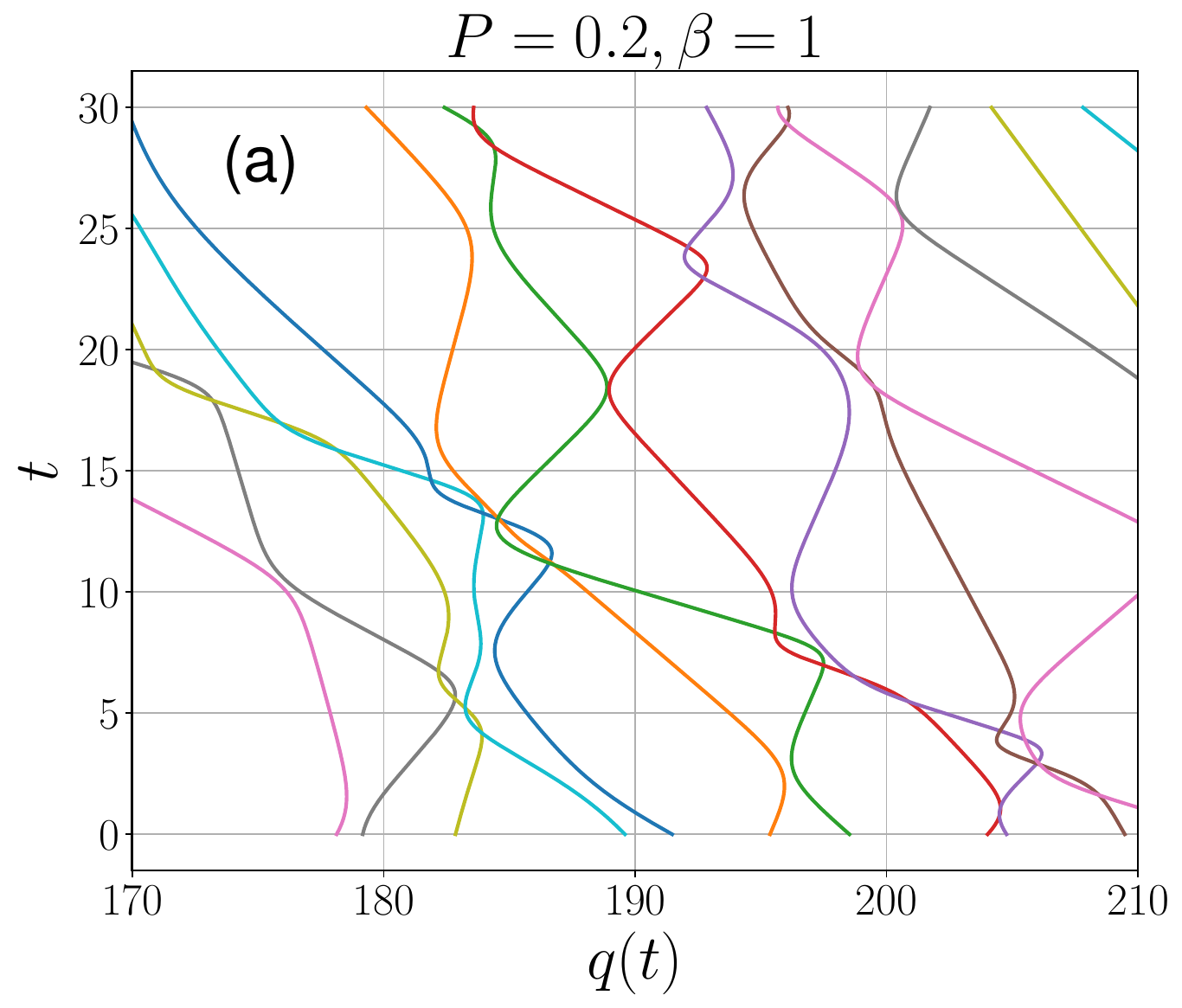}
    \includegraphics[width=7.2cm, height=6.2cm]{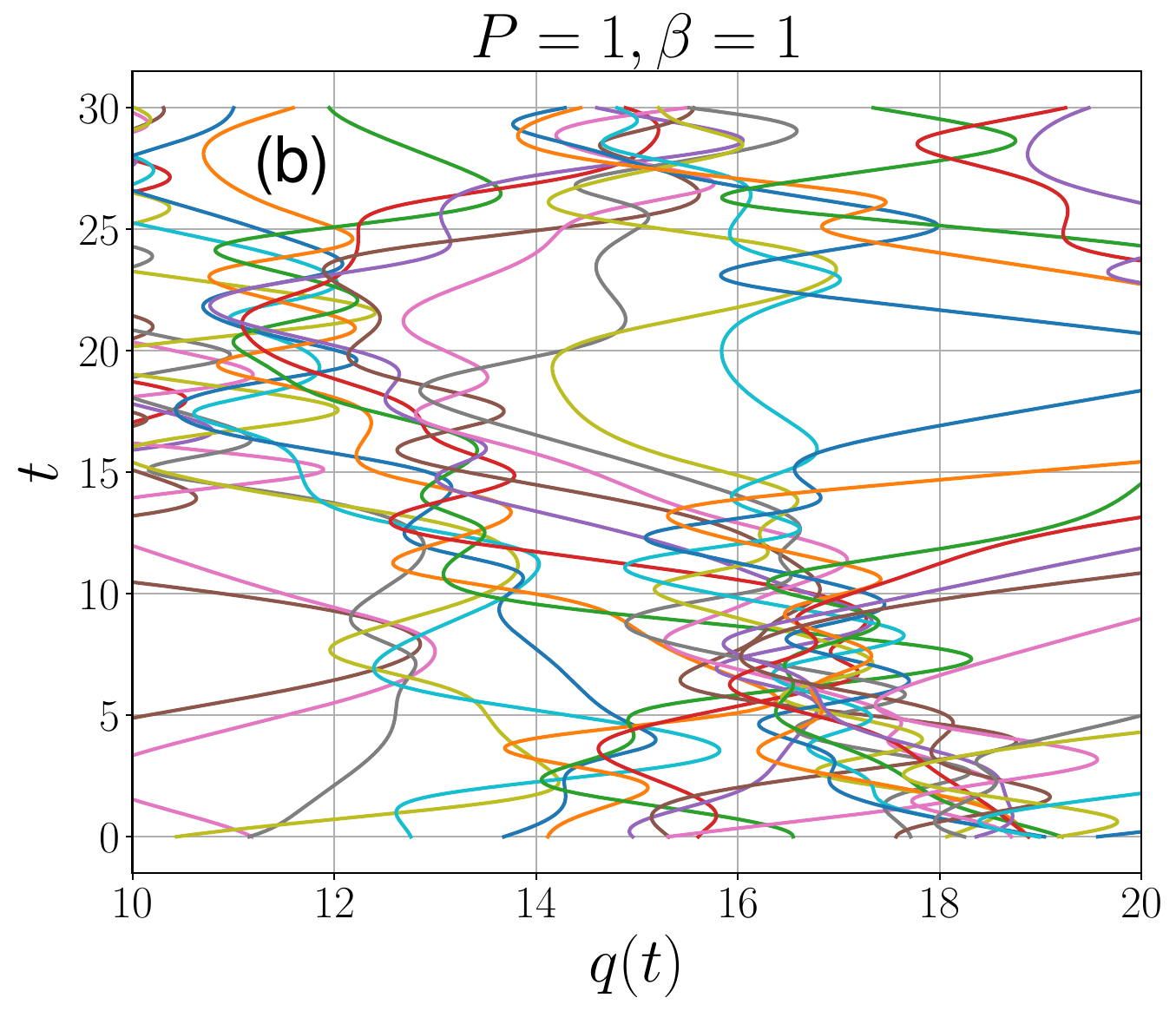}
    \caption{The plots show one sample of Toda particle trajectories. The temperature of the Toda fluid is $1$ and density is $0.25$ in (a) and $1.65$ in (b). For better visibility, these panels are zoomed-in views of trajectories from 100 Toda particles.}
    \label{fig:q(t)}
\end{figure}

In our set-up we prepare the Toda fluid in a thermal equilibrium state. In the framework of generalized hydrodynamics (GHD), integrable many-body systems are viewed as a collection of quasiparticles with velocity dependent scattering shifts. Hence attention has been focused exclusively on the statistical properties of a tagged quasiparticle. Conceptually a quasiparticle 
is defined through approximately maintaining its velocity. As can be seen in Fig. \ref{fig:q(t)}a, such prescription works well at low density. But at higher density a quasiparticle can no longer be located visually, see Fig. \ref{fig:q(t)}b. As properly understood only recently \cite{bulchandani19,lepri25,aggarwal25}, 
the correct tag of a quasiparticle is provided by an eigenvalue of the Lax matrix, while its position is related to the respective eigenvector. The quasiparticle has an effective mean velocity which equals the velocity of the underlying fluid. 
In the context of GHD, the diffusion constant of a tagged quasiparticle has been obtained by designing a simple heuristic picture. The tagged quasiparticle collides with other quasiparticles of the fluid, which travel ballistically with their intrinsic rapidity. The randomness in the initial rapidities and locations of quasiparticles is thereby the source of randomness for the tagged particle. As will be explained, using this picture one arrives at an analytic expression for the diffusion constant \cite{gopalakrishnan18}, see also\cite{denardis18,denardis19}. For the Toda fluid in thermal equilibrium, a proof for the convergence to Brownian motion has been posted a few weeks ago \cite{aggarwal26}, a monumental treatise of close to two hundred pages. 

Besides the standard quasiparticle diffusion, the Toda fluid allows us to explore two further natural options. One is the random motion of eigenvectors. These are localized in the space of particle indices. However, the eigenvectors evolve in time and a natural question concerns their average velocity and its random fluctuations. We derive their effective velocity and we show that they perform a Brownian motion with constant drift. We then compare the numerically obtained diffusion constant with the formula provided to us by Amol Aggarwal, finding excellent agreement.

In addition, also the motion of a tagged particle can be explored. Here, the mean velocity vanishes by symmetry. The particle bounces back and forth between its two neighbors, which themselves have random positions because of thermal fluctuations. As to be discussed, the fluctuations of the tagged particle turn out to be diffusive with an explicit formula for the diffusion constant. Using heuristic arguments based on GHD, we obtain for the velocity auto-correlation the same power law decay as derived by Lebowitz and Percus \cite{lebowitz68} for a fluid of hard rods.
Originally GHD was developed for the time evolution of the densities of conserved fields. The tracking of a specific particle is a novel application.

Tagged quasiparticle motion has  been studied recently also for the box-ball 
system \cite{olla24}. In this model one considers particles located on the one-dimensional lattice $\mathbb{Z}$ with sites either occupied or empty. Time is also discrete. For a single time step, one starts from the very left and consecutively ends at the very right.  As update rules, every particle jumps exactly once and at every occupied site the respective particle jumps to the empty site closest  to its right. Quasiparticles are now solitons, an isolated $n$-soliton consisting of a block of $n$ particles with no holes. As proved in \cite{olla24}, the tagged soliton has an effective velocity and fluctuations governed by Brownian motion. The effective velocity 
and diffusion constant have been computed. As a surprising recurrent feature, while the microscopic integrable many-body systems are diverse, the  hydrodynamic properties
are structurally the same. 

The paper is organized as follows. We first recall properties of the Toda fluid and its thermodynamics in Sec. \ref{sec2}. In Sec. \ref{sec3} we introduce the notion of quasiparticles and show that their mean velocity equals the effective velocity for the conserved densities. In Sec. \ref{sec4}, we derive the diffusion constant under a general GGE for a tagged quasiparticle and in Sec. \ref{sec5} we confirm
the theoretical predictions by molecular dynamics. For this purpose, we generate thermal equilibrium for three distinct values of temperature and pressure. Of course, numerical simulations follow the entire time history for each sample of initial conditions. Thereby we can investigate how fast and how well the expected long time behavior is approached. In  Sec. \ref{sec6} we present similar results for the eigenvectors diffusion. Finally, in Sec. \ref{sec7} we discuss the diffusive behavior of a tagged Toda particle, both proposing a novel derivation for the diffusion constant and the scaling of the velocity auto-correlation function. The predictions are compared to numerical simulations.

\section{Toda fluid}
\label{sec2}
\textbf{Toda dynamics}. The Toda fluid  consists of particles on the real line, position $q_j$ and momentum $p_j$, with exponential  interaction potential governed by the  Hamiltonian \cite{Toda81,spohn24}
\be\label{eq_todahamiltonian}
H_\text{Toda}=\sum_j \Big( \tfrac{1}{2} p_j^2 + e^{-(q_{j+1}-q_j)}\Big).
\ee
The equations of motion read then
\be\label{eq_eom}
\dot{q}_j=p_j, \qquad \dot{p_j} = e^{-(q_j - q_{j-1})} - e^{-(q_{j+1} - q_j)}. 
\ee
We use units such that particle mass and decay scale of the potential are equal to $1$. The interaction is through the \emph{stretch} $r_j=q_{j+1}-q_j$. Since particles interact only through nearest neighbor in index space, one  refers to \eqref{eq_todahamiltonian} also as a chain. Our use of ``fluid'' should emphasize that particles move on the real line. 
Toda knew already about KdV solitons and searched a corresponding feature for discrete nonlinear wave equations. He discovered the exponential chain \cite{toda67} and conjectured integrability, which was then first proved independently by Henon \cite{henon74} and Flaschka \cite{flaschka74}.

We consider a finite number, $N$, of particles with label $j = 1,\dots, N$. Two integrable boundary conditions are widely used.
For the \emph{open} system particles can escape to infinity and the sum in the potential runs over $j = 1,\dots,N-1$, while for the \emph{closed} system
one requires $q_{N+1} = q_1 +\ell$, $\ell$ being the size of the ring, thus controlling the particle density.

In both cases, the Toda fluid is completely integrable. Most directly this can be understood by introducing the \emph{Flaschka variables} 
\be \label{eq:aj_defn}
a_j = \tfrac12\mathrm{e}^{-r_j/2}, 
\ee
in terms of which  the equations of motion in \eqref{eq_eom} read
\begin{equation}\label{eq_eomF}
\frac{d}{dt} a_j = \dfrac{1}{2}a_j(p_j - p_{j+1}),\qquad \frac{d}{dt} p_j = 4(a_{j-1}^2 - a_{j}^2).
\end{equation}
Here $j = 1,\dots, N$ and for the closed fluid the boundary conditions are $a_0 = a_N$, $p_{N+1} = p_1$, while for the open system 
$a_0 = 0 =a_N$.
For the closed system we note that 
\begin{equation}\label{eq_stretch} 
-\frac{d}{dt}\sum_{j=1}^N 2\log(2a_j(t)) = \frac{d}{dt}\sum_{j=1}^N r_j(t) = 0.
\end{equation}
The ring size $\ell$ is conserved, but fluctuates over different  initial conditions. 

Complete integrability of Eq.~\eqref{eq_eomF}  is proved by introducing the Lax matrix $L_N$ and its partner matrix $B_N$ which, in Toda convention, read
\begin{equation}\label{eq_Lax} 
L_N = 
\begin{pmatrix}
\frac{p_1}{2} & a_1&0  &\cdots&a_N\\
a_1 & \frac{p_2}{2} & a_2 & \ddots&0\\
0&a_2& \frac{p_3}{2}&\ddots&\vdots\\
\vdots&\ddots &\ddots&\ddots&a_{N-1}\\
a_N&0&\cdots &a_{N-1}&\frac{p_N}{2}\\
\end{pmatrix},
\quad B_N = 
\begin{pmatrix}
0 & -a_1& 0 &\cdots& a_N\\
a_1 & 0 &-a_2&\ddots  &0\\
0&a_2 & 0&\ddots&\vdots\\
\vdots&\ddots&\ddots& \ddots&-a_{N-1}\\
- a_N&0&\cdots &a_{N-1}&0\\
\end{pmatrix}.
\end{equation}
For the open system one has to set $a_N =0$. $L_N$ is a symmetric Jacobi matrix, while $B_N$ is antisymmetric. These two matrices define the linear system 
\begin{equation}\label{eq_lin}
L_N \ket{\psi_\alpha} = \lambda_\alpha\ket{\psi_\alpha},\qquad  
\frac{d}{dt}\ket{\psi_\alpha(t)}= B_N\ket{\psi_\alpha(t)},
\end{equation}
$\alpha = 1,...,N$, one equation for space and one for time.
The equations of motion, Eq.~\eqref{eq_eomF},  of the Toda fluid now arises as the compatibility condition 
\begin{equation}\label{eq_zerocurv} 
\frac{d}{dt} L_N(t) = [B_N(t),L_N(t)].
\end{equation}
Since the matrix $B_N$ is antisymmetric, it generates the unitary
\begin{equation}
U_N(t) = \exp\Big[\int_0^t ds ~ B_N(s) \Big],
\end{equation}
understood as time-ordered exponential. Using \eqref{eq_zerocurv},  one concludes that $L_N(0)$ and $L_N(t)$ are isospectral, $L_N(t) = U_N(t)L_N(0)U_N(t)^{\rm T}$. The eigenvalues $\lambda_\alpha$ are constant in time and provide $N$ linearly independent conserved quantities. Of course, an eigenvector $\ket{\psi_\alpha} $ evolves in time as 
\begin{equation}\label{eq_unevo} 
\ket{\psi_\alpha(t)} = U_N(t) \ket{\psi_\alpha(0)}.
\end{equation}
The densities of the locally conserved fields are defined by 
\begin{equation}\label{eq_consch} 
Q^{[0]}_{j} = -2\log 2a_j, \qquad Q^{[n]}_{j} = 2(L^n)_{j,j}, \quad n = 1,2,\dots\,,
\end{equation}
implying that their time derivative is the difference of current densities. Using \eqref{eq_zerocurv}, these densities are obtained as 
\begin{equation}\label{eq_currdens} 
J^{[0]}_{j} = -p_j, \qquad J^{[n]}_{j} = 2(L^n)_{j,j+1}a_j, \quad n = 1,2,\dots\,.
\end{equation}
Notice that $Q^{[0]}=\sum_j Q^{[0]}_j$ is the total volume, while $Q^{[1]}=\sum_j Q^{[1]}_j$ and $Q^{[2]}=\sum_j Q^{[2]}_j$ are the total momentum and energy, respectively.\\\\
\textbf{Generalized Gibbs ensembles}.
In generalized hydrodynamics the initial data are random and the standard thermal equilibrium has to be extended to a generalized Gibbs ensemble (GGE)\cite{spohn19_arXiv,spohn24}. 
In Flaschka variables the generalized Gibbs ensemble is written as 
\begin{equation}\label{eq_partf} 
\frac{1}{Z_{N}(P,V)} \prod_{j=1}^{N}  \mathrm{d}p_j  \prod_{j=1}^{N}\mathrm{d}a_j \frac{1}{a_j}
(2a_j)^{2P} \exp\!\big(-\mathrm{tr}[V(L_N)]\big).
\end{equation}
For the closed system this ensemble is stationary in time and invariant under spatial shifts. The factor $1/a_j$ is the Jacobian for the transformation from $q_j$ to $a_j$. The thermodynamic parameters of a GGE are the scalar $P$ and the real-valued function $V$.  As dual of $r_j$, physically $P>0$ is the pressure. $V$ is a potential which confines the eigenvalues.
It could be a finite polynomial, as for thermal equilibrium in which case $V(\lambda) = 2\beta\lambda^2$, $\beta$ the inverse temperature. To have a finite partition function one requires a lower bound as $V(\lambda) > c_0 + c_1|\lambda|$ with $c_1 >0$. A  GGE average is denoted by $\langle\cdot\rangle_{P,V}$, in particularly for thermal equilibrium we write $\langle\cdot\rangle_{P,\beta}$.

Eq. \eqref{eq_partf} is the GGE for a fixed number of particles. Often it is more convenient to work with the grand canonical version. Then $N$ is random and has the weight $\exp[\mu N]$ with $\mu$ the chemical potential. 
 
 Under GGE, the Lax matrix $L_N$ is a random matrix. In general, its matrix elements are correlated. Only for thermal equilibrium they are independent random variables. In the context of GHD, the central property of $L_N$ is the empirical density of states (DOS),
 \begin{equation}
\rho_{\mathrm{DOS},N} (\lambda) = \frac{1}{N}  \sum_{j=1}^N \delta(\lambda - \lambda_j),
\end{equation}
which is a random density normalized to $1$. For large $N$ it converges almost surely to a deterministic limit as
\begin{equation}
 \lim_{N\to\infty} \rho_{\mathrm{DOS},N} (\lambda) = \rho_\mathrm{DOS}(\lambda).
\end{equation}
A more detailed exposition is provided in \cite{spohn24}.\\\\
\textbf{TBA formalism}.
In the infinite volume limit the free energy is determined by a variational problem, where the variation is over positive densities. This leads to Euler-Lagrange equations of a very specific structure, which is known as thermodynamic Bethe ansatz equation (TBA). For the theoretical prediction of the diffusivity we will need some identities from TBA. As common practice the densities are written as functions of the rapidities, $w$. For the Toda fluid rapidities and eigenvalues are related by 
\be
w = 2\lambda.
\ee
One starts from the two-particle scattering shift, for the Toda fluid $\varphi(w,w')=2\log|w - w'|$, and defines the integral operator 
\be\label{eq_shiftop}
Tf(w) = 2\int_\mathbb{R}~dw' \log|w-w'|f(w'),
\ee
acting on some function $f$. The Euler-Lagrange equation is then written for the \emph{pseudoenergy} $\varepsilon(w)$ as 
\be\label{eq_tba}
\varepsilon(w) = V(w) - \mu - T\exp(-\varepsilon(w)),
\ee
where $\mu$ is a chemical potential dual to $P$. Having solved the TBA equation,
one obtains the Maxwell-Boltzmann factor
\be\label{eq_fill2}
\vartheta(w) = \exp(-\varepsilon(w)).
\ee
A numerical procedure to solve the TBA equation is provided in Sec.~\ref{sec5}.
It is convenient to also introduce the quasiparticle density
\be\label{eq_dos_rel}
\rho_\mathsf{p}(w) dw  = 
\nu^{-1}\rho_\mathrm{DOS}(\lambda) d\lambda,
\ee
where $\nu$ is the average free volume, $\nu=\langle r_j\rangle_{P,V}$.
Note that the free volume could be negative and so also $\nu$.
These densities are normalized as
\be
\int_\mathbb{R}~dw\vartheta(w) = P, \qquad
\int_\mathbb{R}~dw\rho_\mathsf{p}(w) = \nu^{-1}.
\ee
It can be shown that
\be 
\rho_\mathsf{p}   = \partial_\mu \vartheta,
\ee
and
\be
\vartheta(w) = \frac{\rho_\mathsf{p}(w)}{1 +  T \rho_\mathsf{p}(w) }.
\ee
Hence $\vartheta(w)$ is also called \emph{filling factor}. One can now introduce the  linear \emph{dressing} operator, denoted by a superscript ``dr", acting on a function $f$ as
\be\label{eq_dress}
f\dr(w) = f(w) + 2\int_\mathbb{R}~dw' \log|w-w'|\vartheta(w')f\dr(w') = (1 - T\vartheta)^{-1}f(w),
\ee
and observes that $((1 - T\vartheta)^{-1})^{\rm T} = (1 - \vartheta T)^{-1}$.
Finally, one introduces the \emph{total space density}
\be\label{eq_fill1}
\rho_\mathsf{s}(w) = 1 +  T \rho_\mathsf{p}(w),
\ee
and notes that 
\be\label{eq_dos_tba}
\rho_\mathsf{s}(w) = [1]\dr(w),\qquad
\rho_\mathsf{p}(w) = \vartheta(w)\rho_\mathsf{s}(w) = \vartheta(w)[1]\dr(w).
\ee

The particle density can be used to compute GGE averages of the conserved fields. More precisely, one confines the particles to a volume of size $\ell$ and takes the limit $\ell \to \infty$, keeping $\nu = \ell/N$ fixed.
Then
\be\label{eq_tbach}
\lim_{\ell\to\infty}\frac{1}{\ell}\langle Q^{[n]}\rangle_{P,V,\ell} = 2^{1-n}\int~dw w^n\rho_\mathsf{p}(w) = 2^{1-n} \int~dw [w^n]\dr(w)\vartheta(w).
\ee
As an example, from Eq. \eqref{eq_consch} for the momentum one dresses the linear function, while for the energy the function is $\tfrac{1}{2} w^2$. Taking derivatives of  \eqref{eq_tbach} with respect to $(P,V)$ also truncated correlation functions can be obtained. 

More explicitly, one perturbs as $V_\kappa = V +\kappa Q^{[m]}$ and takes the derivative \cite{doyon20} 
\be\label{eq_corr}
\lim_{\ell\to\infty}\frac{1}{\ell}\langle Q^{[n]}Q^{[m]}\rangle_{P,V,\ell}^\text{c} = \p_{\kappa}\langle Q^{[n]}\rangle_{P,V_\kappa}\big|_{\kappa = 0} = 2^{1-n}2^{1-m}\int~dw [w^n]\dr(w)[w^m]\dr(w)\rho_\mathsf{p}(w).
\ee
\\
\textbf{Effective velocity}. On the Euler scale, the hydrodynamic equations of the Toda fluid govern the spacetime dependent quasiparticle density $\rho_\mathsf{p}(x,t;w)$ as 
\be
\partial_t \rho_\mathsf{p}(x,t;w) +\partial_x\big(v^\mathrm{eff}(x,t;w)\rho_\mathsf{p}(x,t;w)\big) = 0.
\ee
The breakthrough discovery in 2016 \cite{castro16,bertini16} was to figure out the effective velocity $v^\mathrm{eff}(x,t;w)$. It satisfies the linear integral equation  
\be
v^\mathrm{eff}(w) = w + 2\int_\mathbb{R}~dw' \log|w-w'|\rho_\mathsf{p}(w')\big(v^\mathrm{eff}(w')- v^\mathrm{eff}(w)\big). 
\ee
On an intuitive level, the bare quasiparticle rapidity, $w$, is renormalized by collisions from right and left summed over all other quasiparticles.
The $\log$ provides the size of the jump, while the difference $\big(v^\mathrm{eff}(w')- v^\mathrm{eff}(w)\big)$ is the collision rate.
Using the TBA formalism one obtains the more concise expression
\be\label{eq_tbaveff}
v^\mathrm{eff} = \frac{[w]^\mathrm{dr}}{[1]^\mathrm{dr}}.
\ee
\section{Lax microscope}
\label{sec3}
\textbf{Toda quasiparticles}. In a figurative sense, the Lax matrix serves as a sophisticated microscope. For a given microscopic configuration $\{q_j,p_j\}$ restricted to a particular mesoscopic fluid cell, the values of the conserved fields are hidden. But feeding these data into the matrix $L_N$ restricted to the fluid cell and computing its eigenvalues one observes the local density of states. This density is slowly varying over spacetime. Local free volume and local density of states are the slowly varying degrees of freedom on the hydrodynamic scale.
As noted only very recently \cite{bulchandani19,lepri25,aggarwal25},  in addition the Lax matrix carries the information on the motion of quasiparticles.
To explain, we first should discuss this notion. Roughly speaking the quasiparticle maintains its velocity and interacts with other quasiparticles through pair collisions. The example of a perfect system of quasiparticles is the hard rod fluid
with rod length $a$. Rather than following the particle index the quasiparticle maintains its velocity. Therefore it moves in a straight line interrupted by jumps of size $\pm a$. Quasiparticles of a Toda fluid at very low density behave in the same way, only their scattering shift depends on the incoming velocities. 

\begin{figure}[t]
    \centering
     \includegraphics[width=15.4cm, height=6.5cm]{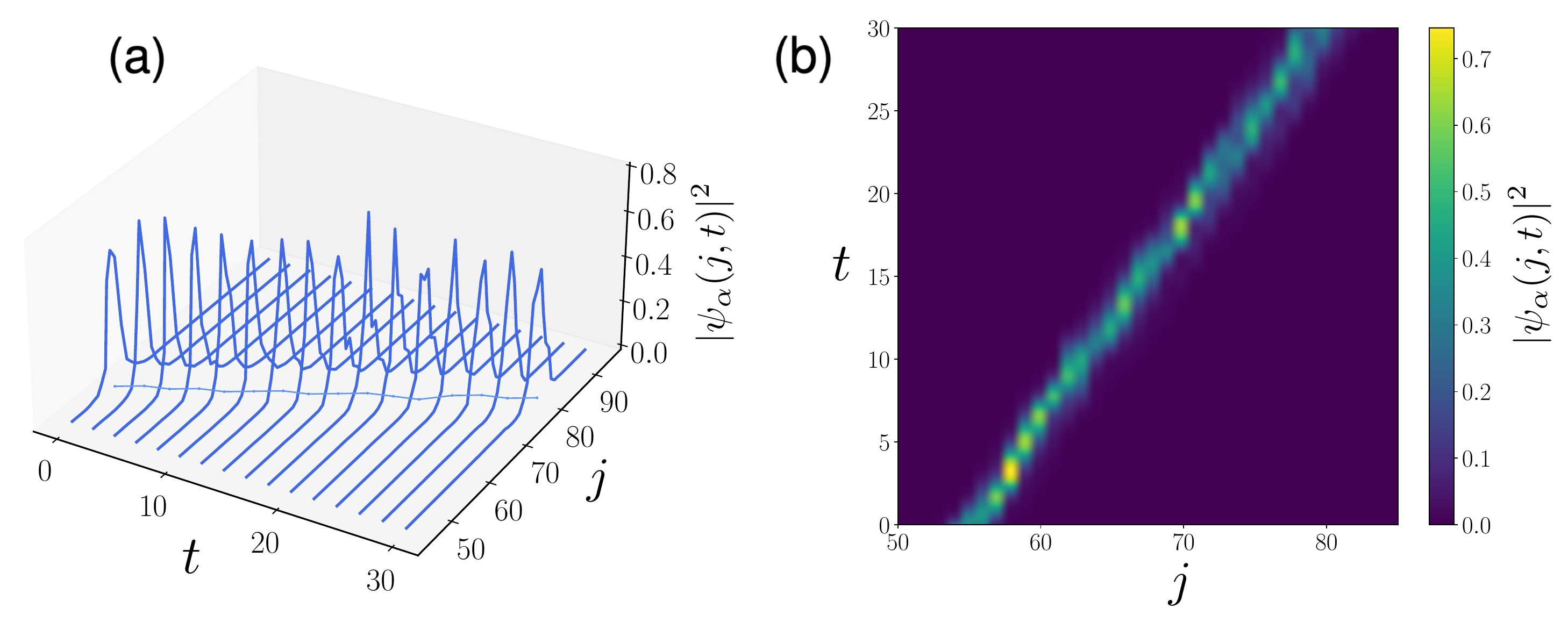}
    \caption{
    Panels (a) and (b) display the evolution of a representative eigenvector corresponding to the eigenvalue $ \lambda_{\alpha} = 0.9976$. In panel (a), the probability density in index space resulting from the eigenvector is shown  at selected time instants. The motion of the peaks is clearly visible and their projection onto the index–time plane is indicated by the light blue line. Panel (b) presents the corresponding heatmap, highlighting the left-to-right propagation of the eigenvector together with its small spatial width. }
    \label{fig:ev-hm}
\end{figure}
\begin{figure}[t]
    \centering
    \includegraphics[width=0.7\linewidth]{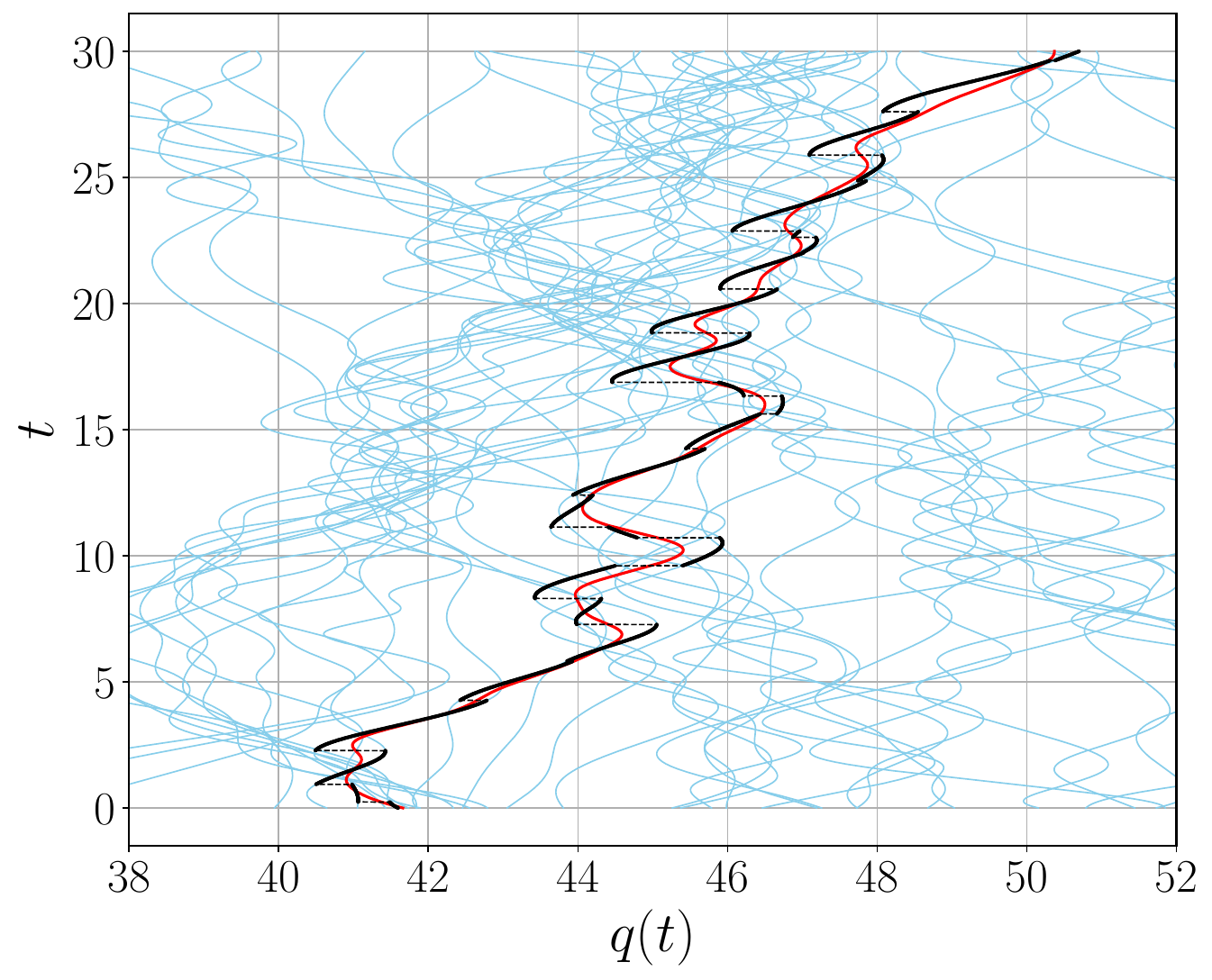}
    \caption{Comparing representative quasiparticle trajectories: the argmax rule (black broken line), having jumps indicated by black dashed segments, and the $|\psi|^2$ rule of Eq.~\eqref{eq_qppos}, yielding a smooth trajectory (red solid line). The quasiparticle is labeled by the eigenvalue $ \lambda_{\alpha} = 0.9976$. The light blue lines display the underlying Toda particle trajectories.}
    \label{fig:Q(t)}
\end{figure}
Surprisingly, even for a very dense fluid, the Lax matrix is able to track the time-dependent position of quasiparticles. They are labeled by the eigenvalues $\lambda_\alpha$ of the Lax matrix
and the position of quasiparticle $\alpha$ is defined by \cite{bulchandani19}
\begin{equation}\label{eq_qppos} 
Q_\alpha(t) = \sum_{j=1}^N |\psi_\alpha(j,t)|^2q_j(t), \quad   \sum_{j=1}^N|\psi_\alpha(j,t)|^2 =1,
\end{equation}
where $\psi_\alpha(j,t)$ is the $j$-th component of the eigenvector. This definition is meaningful, only if the  eigenvector is well localized. For random Jacobi matrices such localization is studied in\cite{teschl2000jacobi}. As discussed in detail in Appendix \ref{appB}, the localization length is related to a strictly positive Lyapunov exponent. Thus
$|\psi_\alpha(j,t)|^2$ is a natural probability distribution for the index of Toda particles on which the quasiparticle  $\alpha$ is sitting. Fig. \ref{fig:ev-hm}a shows how an eigenvector is moving from left to right in index space and the heatmap Fig. \ref{fig:ev-hm}b indicates its width of a few lattice sites. An alternative definition of the quasiparticle position is proposed in \cite{aggarwal25} adopting the following deterministic rule: the  quasiparticle $\alpha$ is located at $q_{j^*}(t)$, where $j^*$ is the argmax of the eigenvector $|\psi_\alpha(j,t)|^2$.   In Fig. \ref{fig:Q(t)} we display both versions. For the argmax rule, there are long jumps, while for the $|\psi|^2$ rule the trajectories change smoothly. By an abstract argument it is proved \cite{aggarwal25} that the precise definition of the position $Q_\alpha(t)$ does not  modify macroscopic quantities, as for example the mean velocity $v^\mathrm{eff}$. 

Introducing the scalar product in $\mathbb{R}^N$ as $\langle \cdot| \cdot \rangle$ and defining the matrix $\mathbb{Q}_N = \mathrm{diag}(q_j)$, Eq. \eqref{eq_qppos} reads more concisely as 
\begin{equation}
\label{2.19} 
Q_\alpha(t) = \bra{\psi_\alpha(t)} \mathbb{Q}(t)\ket{\psi_\alpha(t)}.
\end{equation}
The velocity of the quasiparticle $\alpha$ is then given by 
\begin{align}\label{eq_qpvel} 
\begin{split}
v_\alpha(t) = \frac{\mathrm{d}}{\mathrm{d}t} Q_\alpha(t) =& \sum_{j=1}^N |\psi_\alpha(j,t)|^2p(j,t) + \bra{\psi_\alpha(t)} [B_N(t), \mathbb{Q}_N(t)]\ket{\psi_\alpha(t)}, \\
=& \bra{\psi_\alpha(t)} \mathbb{V}(t)\ket{\psi_\alpha(t)}, 
\end{split}
\end{align}
where we introduced the velocity matrix $\mathbb{V}$ with matrix elements
\begin{equation}\label{eq_velmat} 
\mathbb{V}_{j,j} = p_j, \quad  \mathbb{V}_{j,j+1} = \mathbb{V}_{j+1,j} =r_ja_j, \quad  
\mathbb{V}_{N,1} = \mathbb{V}_{1,N} = r_Na_N,
\end{equation}
and  $\mathbb{V}_{i,j} = 0$ otherwise. Note that the velocity invokes only the eigenvector at time $t$, which can be obtained directly from Eq. \eqref{eq_unevo}. \\\\
\textbf{Average velocity of a tagged quasiparticle}. Physically one expects a quasiparticle to flow with the fluid. In generalized equilibrium the flow is stationary, implying
$\langle v_\alpha(t) \rangle_{P,V}= \langle v_\alpha(0) \rangle_{P,V}$.
Thus one arrives at  the conjecture 
\begin{equation}\label{2.23} 
\langle v_\alpha \rangle_{P,V}
=\langle \bra{\psi_\alpha} \mathbb{V}\ket{\psi_\alpha}\rangle_{P,V} =  v^\mathrm{eff}(\lambda_\alpha),
\end{equation}
valid for a sufficiently large box size $\ell$, where all fields are evaluated at time $t=0$.  

For our choice of $Q_\alpha $, there is a simple argument which verifies the conjecture. 
On the Toda fluid side we start from the continuum version of the local conservation laws.  For the $n$-th charge, the continuum density is given by  
\begin{equation}\label{2.24} 
Q^{[0]}_\mathsf{f}(x) = \sum_{j \in \mathbb{Z}} \delta(q_j - x),\qquad Q^{[n]}_\mathsf{f}(x) = \sum_{j \in \mathbb{Z}} 
\delta (q_j- x) Q^{[n]}_{j-1}.
 \end{equation}
In the latter sum one might have expected the index $j$, but $j-1$ better fits the algebra. 
 The conservation laws read 
 \begin{equation}\label{2.25} 
\partial_t Q^{[n]}_\mathsf{f}(x,t) + \partial_x J^{[n]}_\mathsf{f}(x,t) = 0,\quad n \geq 0.
\end{equation}
Working out the derivatives, compare with \cite{spohn24}, Chapter 9, one obtains the fluid current densities
 \begin{eqnarray}\label{eq_fluidcurrents} 
&&\hspace{0pt}J^{[0]}_\mathsf{f}(x) = \sum_{j \in \mathbb{Z}}\delta(q_j - x)p_j, \\
&&\hspace{0pt} J^{[n]}_\mathsf{f}(x) = \sum_{j \in \mathbb{Z}}\Big(\delta(q_j - x)p_jQ^{[n]}_j + \big(\theta(q_{j} -x) - \theta(q_{j-1} -x)\big)J_{j-1}^{[n]}\Big),\nonumber
\end{eqnarray}
with $\theta$ the step function, $\theta(x) =0$ for $x\leq 0$ and $\theta(x) =1$ for $x > 0$. We now integrate the current over the interval 
$[0,\ell]$ and get  
\begin{equation}\label{2.27}
\int_0^\ell dx J^{[n]}_\mathsf{f}(x) = \sum_{j \in \mathbb{Z}} \mathds{1}([0,\ell])(q_j ) \big(p_j Q^{[n]}_j + 2r_{j-1}a_{j-1} (L^n)_{j-1,j}\big),
\end{equation}
 $\mathds{1}(\{\cdot\})$ being the indicator function for the set $\{\cdot\}$. 
On the side of quasiparticles, we set $\mathsfit{v}(\lambda_\alpha) = v_\alpha$, multiply by $2(\lambda_\alpha)^n$, and sum over $\alpha$. Then we get
\begin{equation}\label{2.28}
\sum_{\alpha=1}^N 2(\lambda_\alpha)^n \mathsfit{v}(\lambda_\alpha) = \sum_{\alpha=1}^N 2(\lambda_\alpha)^n  \bra{\psi_\alpha} \mathbb{V}\ket{\psi_\alpha} = 2\mathrm{tr}(\mathbb{V}(L_N)^n)
= \sum_{j=1}^N\Big(Q^{[n]}_{j}p_j + 2r_ja_j((L_N)^n)_{j,j+1} \Big).
\end{equation}
We now first average the fluid side. As discussed in \cite{spohn24}, Chapter 9, this results in
\begin{equation}\label{2.29} 
\lim_{\ell \to \infty} \frac{1}{\ell} \int_0^\ell dx \langle J^{[n]}_\mathsf{f}(x) \rangle_{P,V} = 2\int d \lambda \rho_{\mathrm{DOS}}(\lambda)v^\mathrm{eff}(\lambda)\lambda^n.
\end{equation}
 On the side of quasiparticles,
\begin{equation}\label{2.30}
\lim_{N \to \infty} \frac{1}{N} \sum_{\alpha=1}^N 2(\lambda_\alpha)^n \mathsfit{v}(\lambda_\alpha) = 2\int d \lambda \rho_\mathrm{DOS}(\lambda) \mathsfit{v}(\lambda)\lambda^n.
\end{equation}
Since the power $n$ is arbitrary, we conclude that 
 \begin{equation}\label{2.31} 
 \mathsfit{v}(\lambda) = v^\mathrm{eff}(\lambda).
\end{equation}

\section{Quasiparticle diffusion, theory} 
\label{sec4}
To compute the diffusion constant of a tagged quasiparticle we exploit a mesoscopic description based on TBA.
This argument has been developed  before \cite{gopalakrishnan18}. 
Since this is the key formula to be compared with numerical simulations,
we supply the entire derivation, specialized to the Toda fluid
which adds some simplifications. 

\begin{figure}[t]
    \centering
    \includegraphics[width=0.75\linewidth]{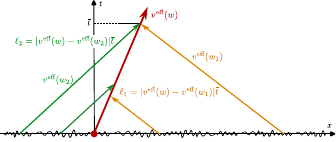}
    \caption{Schematically illustrating the contribution of initial-state noise to diffusion. A tagged quasiparticle, initially at the origin (red line), moves with an effective velocity $v^\text{eff}(w)$ through a noisy, homogeneous background. Quasiparticles from each mesoscopic fluid cell propagate with distinct effective velocities and fluctuating particle numbers. These density fluctuations propagate ballistically and scatter with the tagged quasiparticle. Over a time interval $\bar t$, the tagged quasiparticle trajectory is influenced by quasiparticles of rapidity $w'$ that originated within a spatial region of extent $\ell=|v^\text{eff}(w)-v^\text{eff}(w')|\bar t$.}
    \label{fig:qpdiff}
\end{figure}

The tagged quasiparticle with rapidity $w$ starts from the origin and moves ballistically with effective velocity $v^{\rm eff}(w)$ due to interaction with other quasiparticles in the medium.
Until time $t$, the tagged quasiparticle collides with other quasiparticles traveling with their effective velocity $v^{\rm eff}(w')$  only if they have started within a spatial window of size $\ell_{w,w'} = |v^{\rm eff}(w)-v^{\rm eff}(w')|t$, see Fig. \ref{fig:qpdiff}.

As recently established for integrable systems\cite{doyon26noise}, noise is entirely restricted to initial conditions and propagates ballistically. Summing contributions from all colliding fluid particles, the position of the quasiparticle at time $t$ can be written as
\begin{align}\label{eq:Qwt}
  Q(w,t)  = v^\mathrm{eff}(w)t + t\int dw'   \frac{\delta v^\text{eff}(w)}{\delta \vartheta(w')}
 \overline {\delta \vartheta}(w'),
\end{align}
where $\overline {\delta \vartheta}(w')$ are the space averaged initial time fluctuations
\begin{align}
\overline {\delta \vartheta}(w')=\frac{1}{\ell_{w,w'}}\int_0^{\ell_{w,w'}}dx~\delta \vartheta(w',x).
\end{align} 
Because the system features rapid decay of correlations, for $\ell_{w,w'}$ being much larger than the correlation length, typical fluctuations $\overline {\delta \vartheta}(w')$ are of order $1/\sqrt{\ell_{w,w'}}$.

To obtain the average of $\big(Q(w,t) - v^\mathrm{eff}(w)t\big)^2$, we have to figure out the variational derivative and also the GGE average $\langle \overline {\delta \vartheta}(w')\overline {\delta \vartheta}(w'') \rangle_{P,V}$. 

To compute the first item, we apply simple algebra to the dressing definition in Eq. \eqref{eq_dress} to show that $\delta f\dr(w)/\delta \vartheta(w')=\varphi\dr(w,w')f\dr(w')$, where $\varphi\dr$ is the dressed scattering shift. Using Eq. \eqref{eq_tbaveff}, this yields
\begin{equation}
\label{3.2}
\delta v^\mathrm{eff}(w)/\delta \vartheta(w') = \frac{1}{\rho_\mathsf{s}(w)}\varphi^\mathrm{dr}(w,w')\rho_\mathsf{s}(w')(v^\mathrm{eff}(w')-v^\mathrm{eff}(w)).
\end{equation}
To compute $\langle \overline {\delta \vartheta}(w')\overline {\delta \vartheta}(w'') \rangle_{P,V}$, we first note 
from Eq. \eqref{eq_fill1} that $\delta\vartheta(w,x)=\rho_\mathsf{s}^{-1}(w)(1-\vartheta T)\delta\rho_\mathsf{p}(w,x)$. Now using [see \cite{doyon20}, Section 3.5]
\begin{align}\label{eq:delrho_p}
\langle\delta\rho_\mathsf{p}(w',x)\delta\rho_\mathsf{p}(w'',y)\rangle_{P,V}=(1-\vartheta T)^{-1}\rho_\mathsf{p}\delta(w'-w'')\delta(x-y)(1-T\vartheta)^{-1},
\end{align}
one finds (also see the arguments given in \cite{gopalakrishnan18} )
\begin{align}
\langle \overline {\delta \vartheta}(w')\overline {\delta \vartheta}(w'') \rangle_{P,V}=\frac{1}{\ell_{w',w''}}\delta(w'-w'')\vartheta(w')\rho_\mathsf{s}^{-1}(w'). \label{eq:bthbth_corr}
\end{align}
Using the results from Eqs.~\eqref{3.2} and \eqref{eq:bthbth_corr}, one arrives at  
\begin{equation}
\label{eq:mfrk(D)}
\big(Q(w,t) - v^\mathrm{eff}(w)t\big)^2 = t 
 \int~dw' \rho_\mathsf{p}(w')\left( \frac{\varphi^\mathrm{dr}(w,w')}{\rho_\mathsf{s}(w)} \right)^2|v^\mathrm{eff}(w)-v^\mathrm{eff}(w')| := 2 \mathfrak{D}(w)t.
\end{equation}
$ \mathfrak{D}(w)$ is the diffusion coefficient of the tagged quasiparticle with rapidity $w$.

Under GGE, on a somewhat coarser spatial scale, the initial fluctuations are Gaussian white noise in space. Going back to Eq. \eqref{eq:Qwt}, this means that the quasiparticle  is driven by white noise, implying that $(Q(w,t)
- v^\mathrm{eff}(w)t)/\sqrt{t}$ has the statistics of a Brownian motion with diffusion constant  $\mathfrak{D}(w)$.
Remarkably, the same formula has been found in Refs. \cite{denardis18,denardis19} by merely considering the diagonal part of the bulk diffusion matrix obtained from the Onsager matrix, see Appendix \ref{appA} for more explanations.

The identity \eqref{eq:mfrk(D)} for $\mathfrak{D}(w)$  can be numerically evaluated by solving the TBA equations, see Sec. \ref{sec5} for details. Fig. \ref{fig:tbadiff} displays some examples for thermal states at $\beta=1$ and varying pressure. 

\section{Quasiparticle diffusion, molecular dynamics}\label{sec5}
In this section, we report on numerical calculations and compare with theoretical results.
The first step is to sample initial conditions for $N$ Toda particles, $\{q_j(0), p_j(0)\}$, according to thermal equilibrium. Then $V(\lambda) = 2\beta \lambda^2$. According to \eqref{eq_partf}, the $\{p_j\}$ are i.i.d. Gaussian random variables with variance $1/\beta$. The $\{a_j\}$ are i.i.d. $\chi$ distributed random variables. To facilitate sampling, it is convenient to switch from $\chi$ to the Gamma distribution by the change of variables $a_j^2 = u_j$. Then the $u_j$'s are distributed according to  the probability density 
\begin{equation}
\label{5.1}
\frac{(4\beta)^P}{\Gamma(P)} u^{P-1}e^{-4\beta u}\theta(u) du.
\end{equation}
Sampling the $u_j$'s, one obtains the stretch variables $r_j= - \log (4 u_j)$ and the positions by using  $q_{j+1}=q_j+r_j$ upon setting $q_1=0$. For the \emph{closed} system, $q_{N+1}=q_1 + \ell$, where the ring size $\ell$ varies from one realization to another. The equilibrium state is characterized by the average stretch per particle, $\nu = \langle \ell \rangle_{P,\beta}/N$.

Once the entire initial configuration $\{q_j,p_j;~j=1,2,\dots,N\}$ is chosen, the Lax matrices $(L_N,B_N)$ can be constructed. The diagonalizing $L_N$ yields the eigenvalues $\lambda_\alpha$ and the corresponding eigenvectors $\ket{\psi_\alpha}$. 
As Toda particles evolve according to Eq.~\eqref{eq_eom}, the Lax matrix evolves as well. Its eigenvalues remain constant in time while  the eigenvectors are governed by  Eq.~\eqref{eq_lin} as 
\begin{align}\label{eq_lin-sec5} 
\begin{split}
 & \hspace{3cm} \frac{d}{dt}{\psi}_\alpha(j,t) = \sum_{i = 1}^N (B_{N}(t))_{j,i}\,\psi_\alpha(i,t)\, , \\
& (B_N)_{j+1,j} = -(B_N)_{j,j+1} = \tfrac{1}{2}e^{-r_j/2}\, ,~~~ (B_N)_{1,N}=-(B_N)_{N,1}=\tfrac{1}{2}e^{-r_N/2}.
\end{split}
\end{align}

The equations of motion~\eqref{eq_eom} for the particle coordinates $(q_j, p_j)$ and the corresponding evolution equations for the eigenvectors in Eq.~\eqref{eq_lin-sec5} are integrated using the standard fourth-order Runge-Kutta (RK4) scheme. 
Particle trajectories are shown in Fig.~\ref{fig:q(t)}, whereas Fig.~\ref{fig:ev-hm} displays the evolution of an eigenvector. 
\\\\
\textbf{Numerical implementation and results}.\label{sec:numerics_binning}
The eigenvalues of the Lax matrix $L_N$ depend on initial configurations, which are distributed according to the thermal equilibrium ensemble.  To condition this ensemble on a specific eigenvalue seems to be very difficult, both analytically and also numerically.
Hence to compute quasiparticle properties as a statistical average over initial configurations requires a spectrally resolved averaging. For this purpose we first discretize spectral space into bins of uniform width, $\Delta$, covering the finite interval $[w_{\rm min},w_{\rm max}]$. Each bin $\mathsfit{b}$ is centered at $w_{\mathsfit{b}}$ and corresponds to the interval $[w_{\mathsfit{b}}-\tfrac{1}{2}\Delta,w_{\mathsfit{b}}+\tfrac{1}{2}\Delta]$. 

Before explaining  the sampling scheme, it is convenient to slightly abstract from our concrete set up.
Given is the $N$-particle phase space, $\Gamma$. The points of $\Gamma$ are denoted by $y =(q,p)$. Then there is the collection of rapidities, $w_{\alpha} $, and the observable $\mathcal{O} = \{\mathcal{O}_1,\dots,\mathcal{O}_N\}$, all as functions on $\Gamma$. Furthermore, given is the thermal ensemble
$\rho_\mathrm{th}(y)dy$. We now fix some bin $\mathsfit{b}$ and want to compute the average 
of $\mathcal{O}$ under the constraint $w_\alpha \in \mathsfit{b}$. This average is denoted by $\langle \mathcal{O}|| \mathsfit{b} \rangle_{\mathrm{th}}$ and the averaging prescription reads
\be\label{eq_y}
\langle \mathcal{O}||\mathsfit{b}   \rangle_{\mathrm{th}} 
 = \frac{\int dy \rho_\mathrm{th}(y) \big(\sum_{\alpha = 1}^N \mathds{1}(\{w_\alpha(y) \in \mathsfit{b}\})\mathcal{O}_\alpha(y)\big)}
 {\int dy \rho_\mathrm{th}(y) \big(\sum_{\alpha = 1}^N\mathds{1}(\{w_\alpha(y) \in \mathsfit{b}\})\big)}. 
 \ee
 The subscript refers to the thermal average. Later on we will also use the subscript $(P,\beta,N)$ and $(P,\beta) $ for infinite volume. In the numerator, $w_\alpha$ has to lie in the bin $\mathsfit{b}$ and averaged is the $\alpha$-component of $\mathcal{O}$.
The denominator arises from normalizing so that $\langle 1||\mathsfit{b}   \rangle_{\mathrm{th}} = 1$.
Sampling will be separate for numerator and denominator and amounts to substituting the $y$-integral by a sum over $\mathcal{R}$ independent realizations. In practice we have $N=500$ and choose $\Delta$ so small that there are 240 bins covering the interval on which the density of states is concentrated.

\begin{figure}[h]
    \centering
    \includegraphics[width=4.9cm, height=5.0cm]{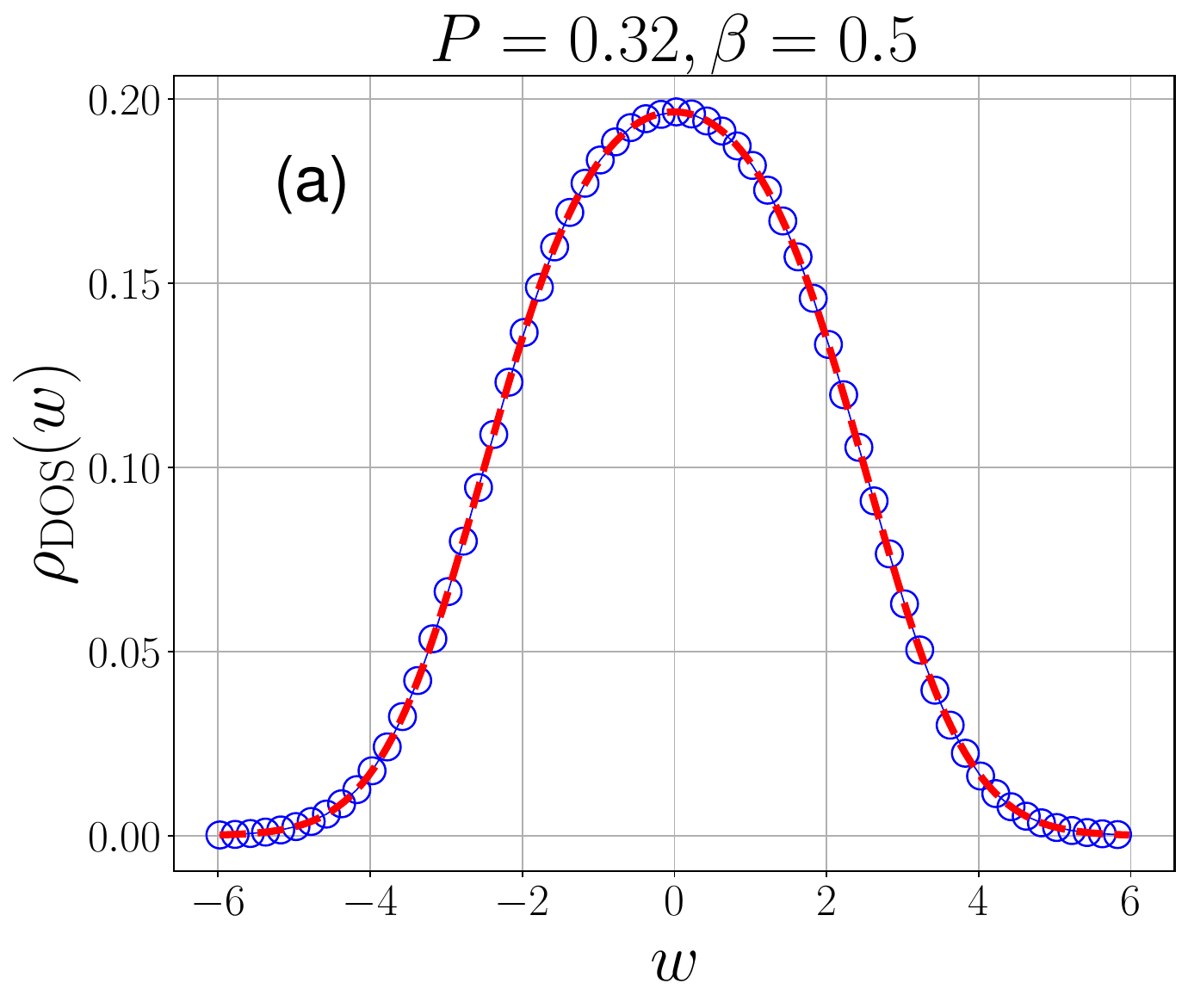}
    \includegraphics[width=4.9cm, height=5.0cm]{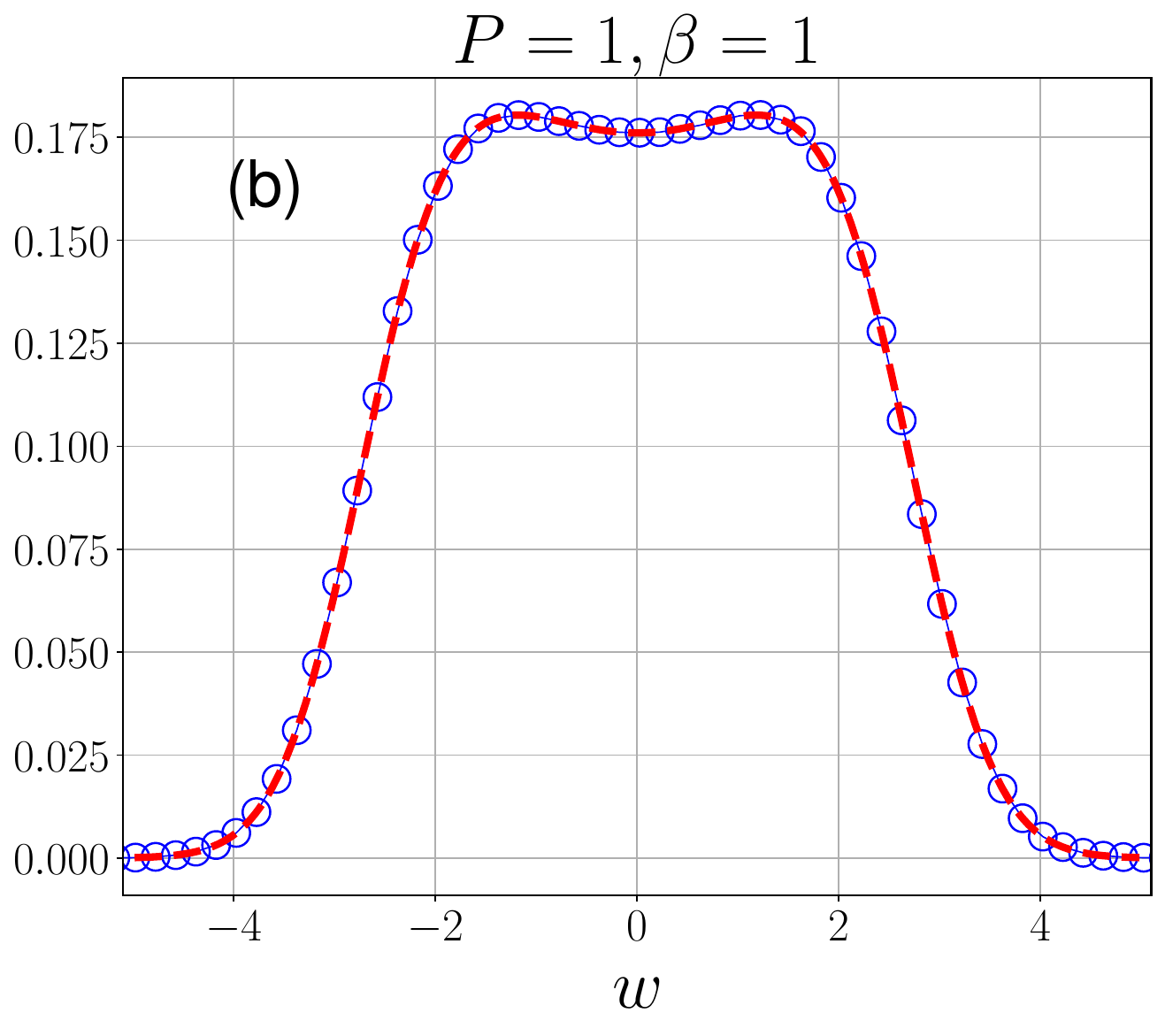}
    \includegraphics[width=4.9cm, height=5.0cm]{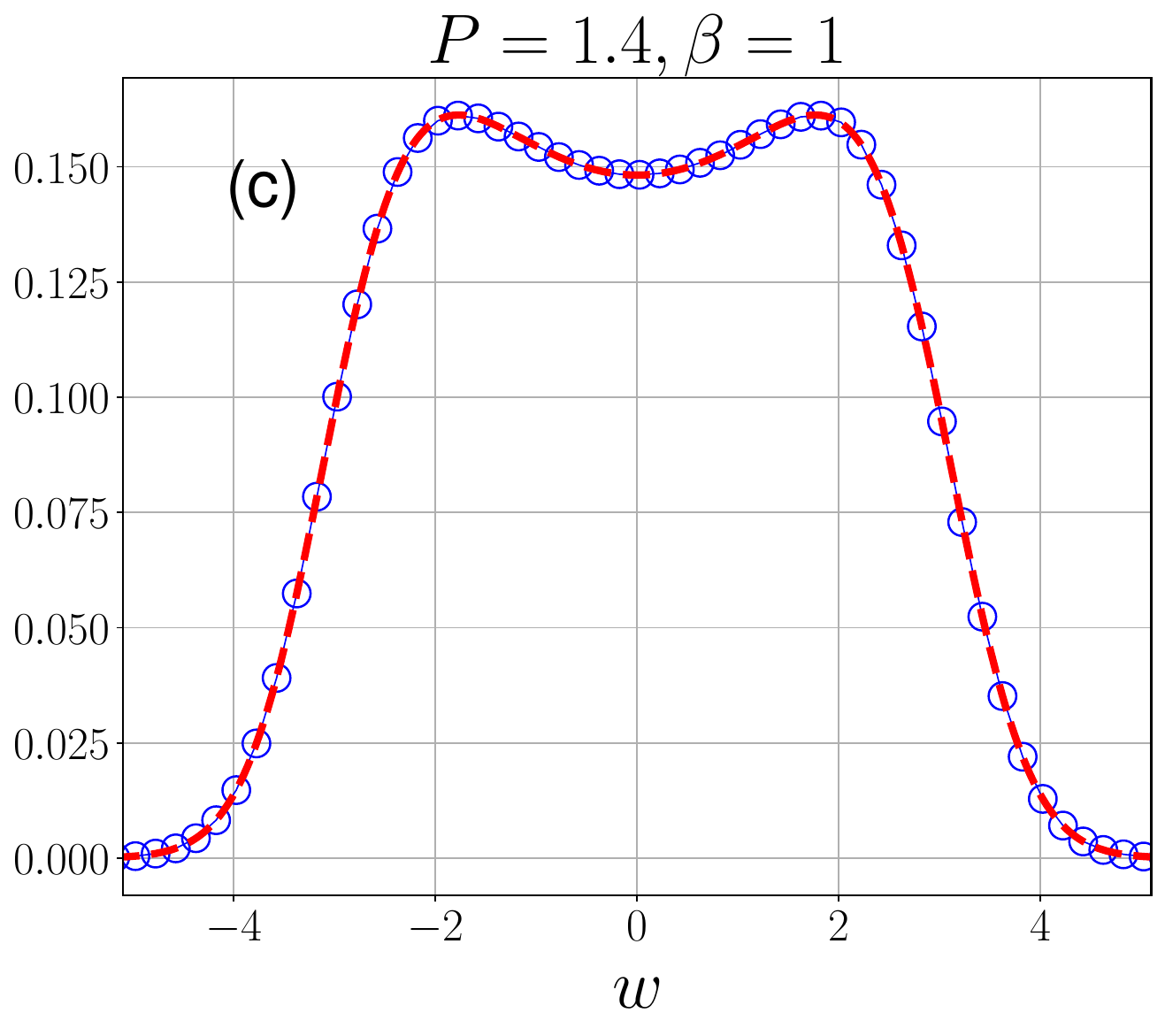}\caption{Plots comparing the simulation (blue circles) and analytical results given by Eq.~\eqref{eq_dos_rel} (red dashed line) for the density of states across three set of parameters: (a) $P=0.32,~\beta=0.5,$ yielding an average stretch per particle $\nu = 2.58$, (b) $P=1,~\beta=1,$ with $\nu = 0.58,$ and (c)   $P=1.4,~\beta=1,$ with $\nu = 0.06$. For all three plots, $N=500$, $\Delta=0.05$, and averages have been taken over $\mathcal{R} = 10^6$ samples. The density of the Toda fluid is increasing from left to right.}
    \label{fig:DOS}
\end{figure}

The spectral density equals the denominator 
in Eq.~\eqref{eq_y}. It can be written more concisely in terms of the occupations $N_{\mathsfit{b}}$ counting the number of eigenvalues in bin $\mathsfit{b}$. Adding the volume element $1/\Delta$ and the numerical sum over realizations, one arrives at the expression 
\be 
\frac{1}{\Delta} \frac{1}{\mathcal{R}}\sum_{{\mathtt r}=1}^{\mathcal R} N_{\mathsfit{b}}(y_{\mathtt r}),
\ee
which  should converge to the thermal equilibrium $\rho_\mathrm{DOS}$ for large $N,\mathcal{R}$ and sufficiently small $\Delta$. This expectation is well confirmed numerically, see Fig. \ref{fig:DOS}.
In all three parameter sets, the MD simulations show excellent  agreement with the theoretical predictions from TBA solutions, see end of this section. 
\begin{figure}[t]
    \centering
    \includegraphics[width=4.9cm, height=5.0cm]{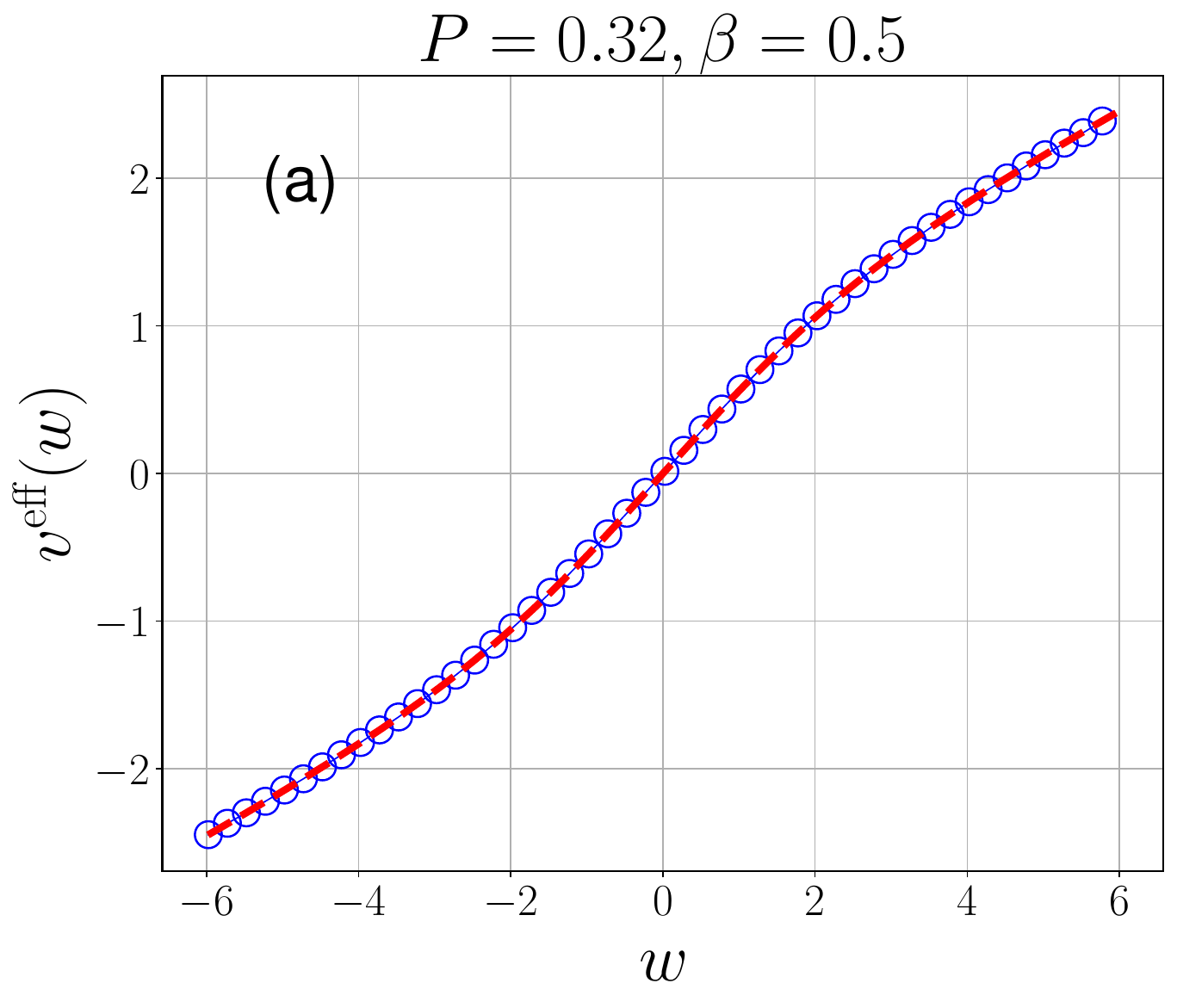}
    \includegraphics[width=4.9cm, height=5.0cm]{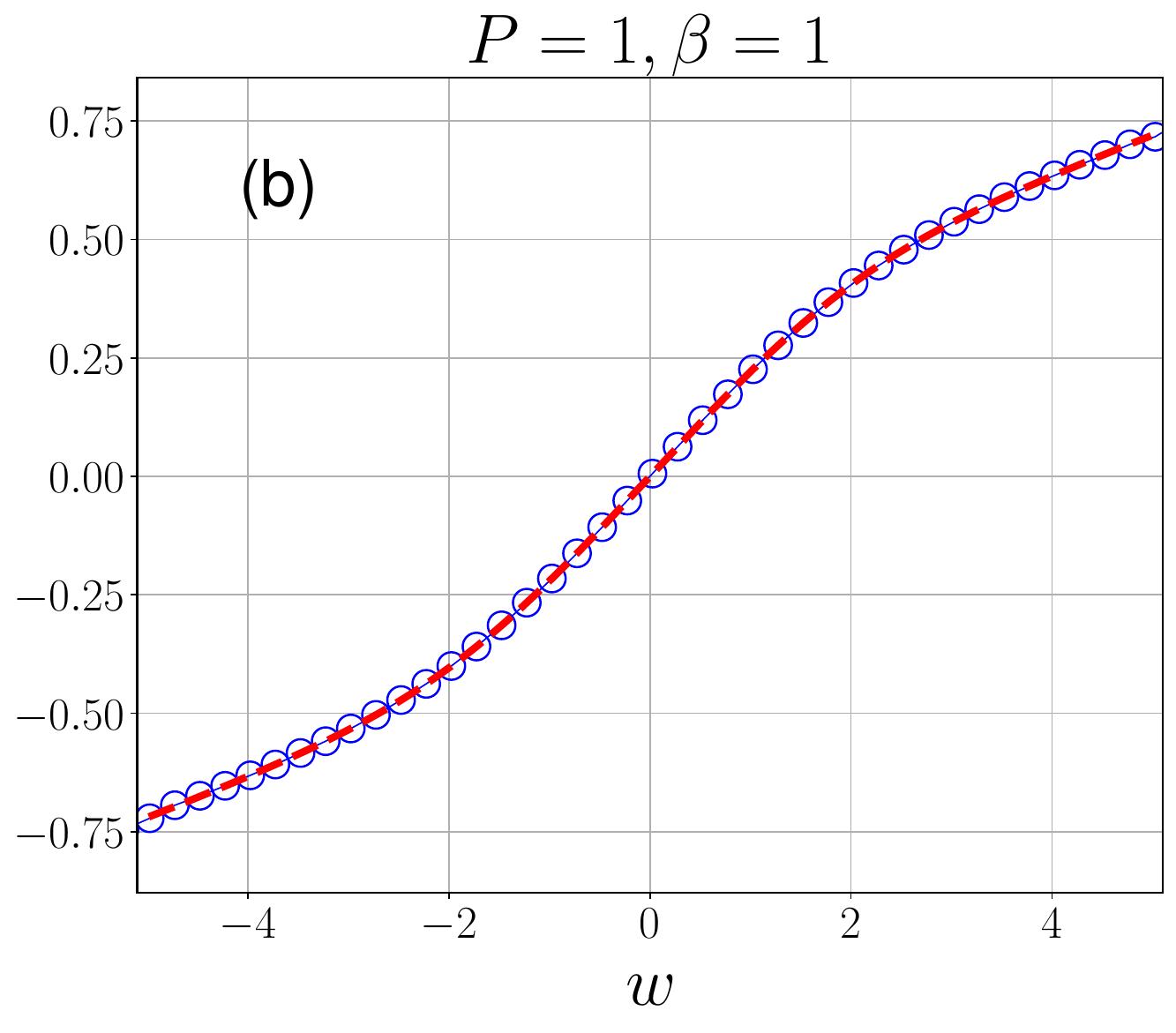}
    \includegraphics[width=4.9cm, height=5.0cm]{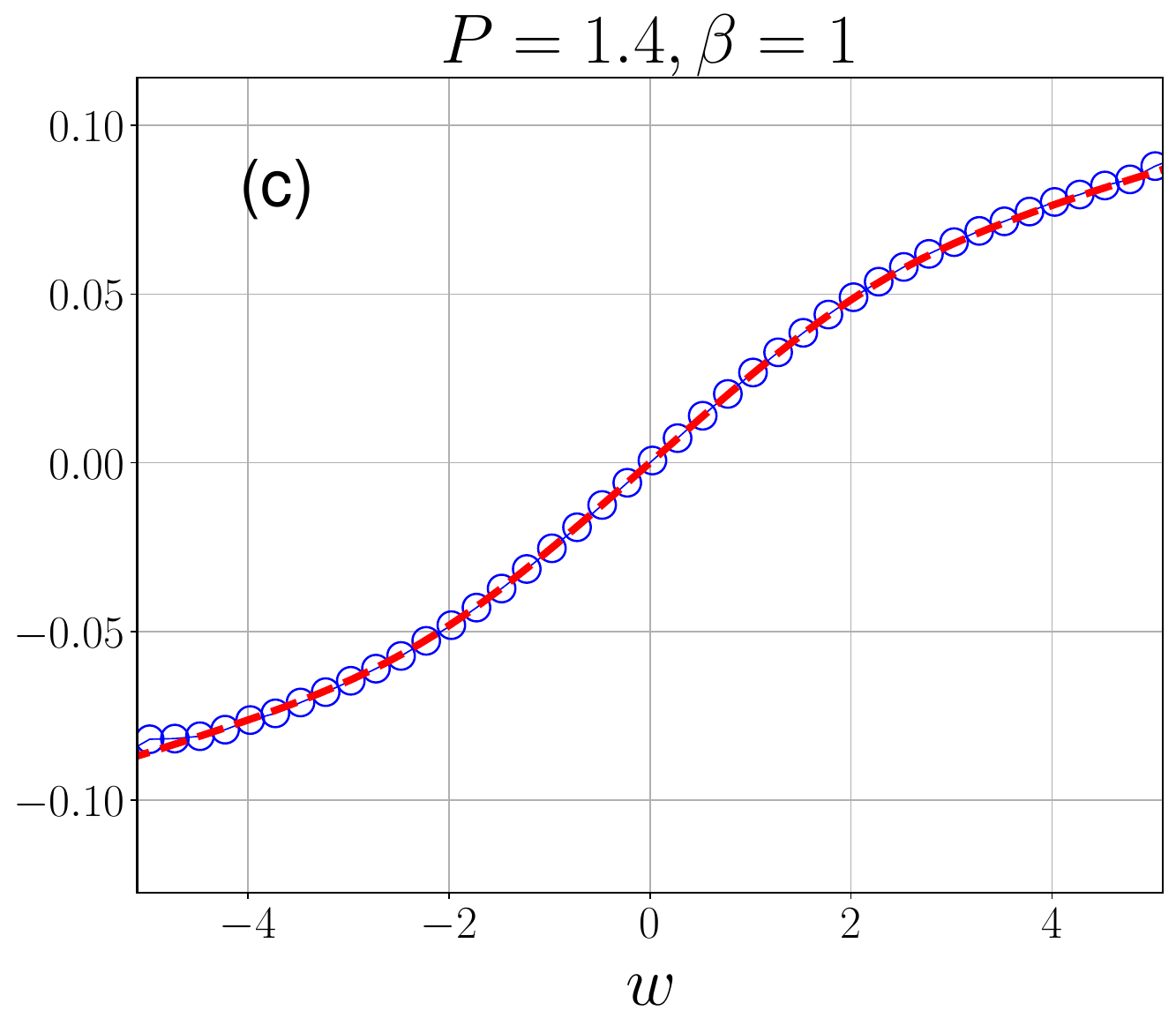}
    \caption{Comparison of the simulated effective quasiparticle velocity $v^\mathrm{eff}(w)$ (blue circles) with the analytical prediction from Eq.~\eqref{eq_tbaveff} (red dashed line) for the same parameter sets as in Fig.~\ref{fig:DOS}. }
    \label{fig:Eff_velocity}
\end{figure}
At low pressure, the system is dilute and the DOS takes a Gaussian 
shape, since particle interactions are weak and the spectrum is dominated by the
thermal Maxwellian. At intermediate pressure, interactions 
become more significant and the DOS develops a nearly flat-top. At $P = 1.46, \beta = 1$, the stretch vanishes and the particle density diverges. However DOS, $v^\mathrm{eff}$, and 
$\mathfrak{D}(w)$ seem to change smoothly through this parameter point. For even larger pressure the two peaks become much more pronounced and the outer slopes decay extremely rapidly \cite{kethepalli25, mendl22}. In the diagonal limit  
$\beta \to \infty$ and $P \to \infty$,  keeping $\mathfrak{p}=P/\beta$ finite, the DOS converges to the arcsine distribution \cite{spohn24}.

As further task we simulate the average velocity, which amounts to the choice $\mathcal{O}_\alpha= v_\alpha$.
Our numerical results are shown in Fig.~\ref{fig:Eff_velocity}. For all three parameter sets, the MD results agree precisely with the TBA predictions Eq.~\eqref{eq_tbaveff}. We observe that the effective velocity has a characteristic S-shape, being essentially linear near $w=0$.  At very low  pressure the slope near $0$ would be $1$, but our lowest pressure shows already slope $1/2$ as a result of the interaction with other quasipaticles. This effect becomes even more pronounced at higher pressures.

Most importantly we arrive at the observable of key interest, namely the velocity auto-correlation function which is the constrained average
of $\mathcal{O}_\alpha = v_\alpha(t)v_\alpha(0)$. As $t \to \infty$ this average converges to $(v^\mathrm{eff})^2$. 
Thus one switches to the truncated velocity auto-correlation function,
\be\label{eq:vcorr_quasi}
C_v(w,t) = \langle v_\alpha(t)v_\alpha(0)||\mathsfit{b}_w    \rangle_{P,\beta}
- v^\mathrm{eff}(w)^2,
\ee 
where $\mathsfit{b}_w$ is the bin containing $w$. The time-dependent diffusion constant is then defined as 
\be\label{D(w,t)-qp}
\mathfrak{D}(w,t) = \int_0^t ds \, C_v(w,s).
\ee
Its long time limit is the diffusion constant $\mathfrak{D}(w)$.

\begin{figure}[h]
    \centering
    \includegraphics[width=4.9cm, height=5.0cm]{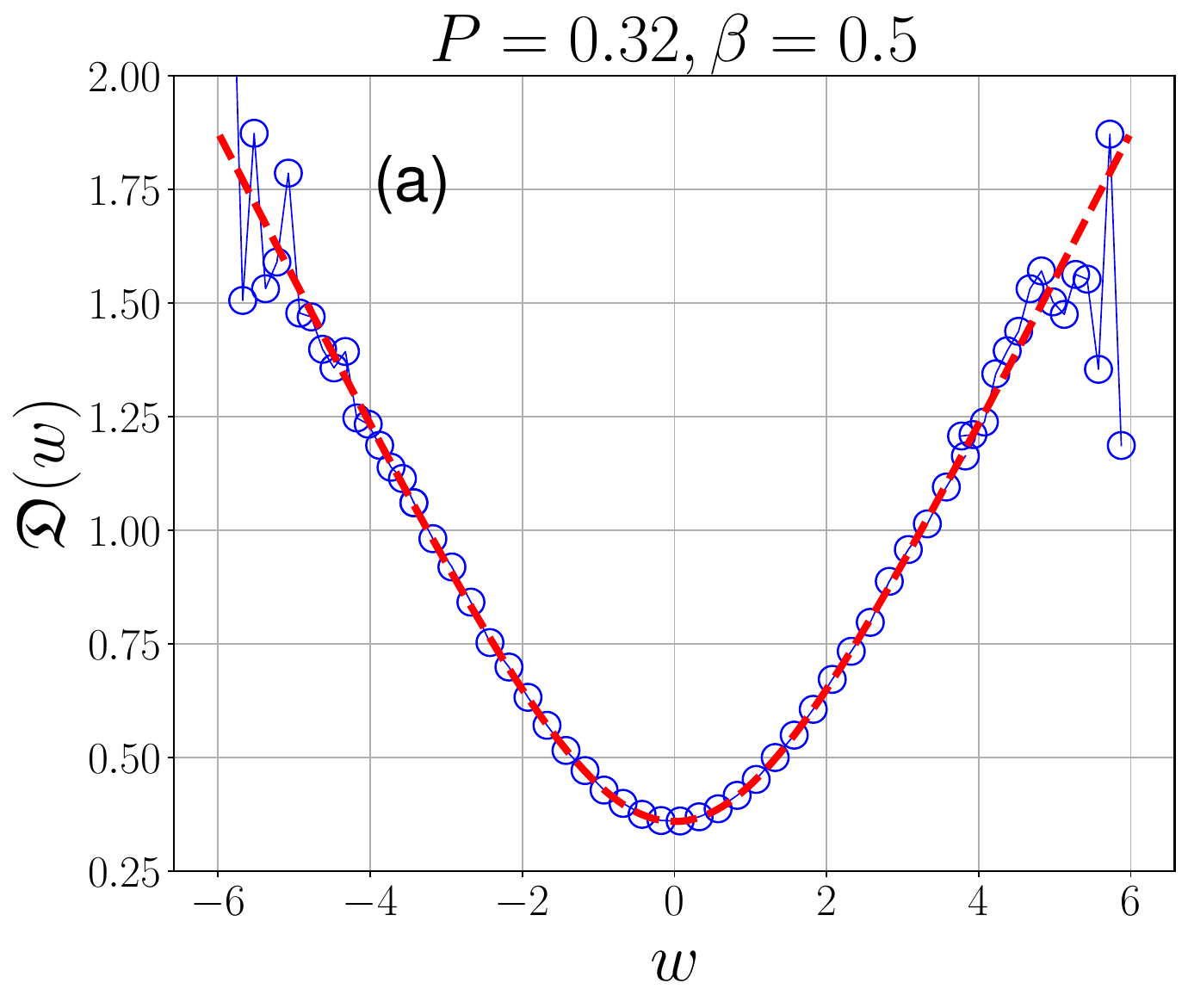}
    \includegraphics[width=4.9cm, height=5.0cm]{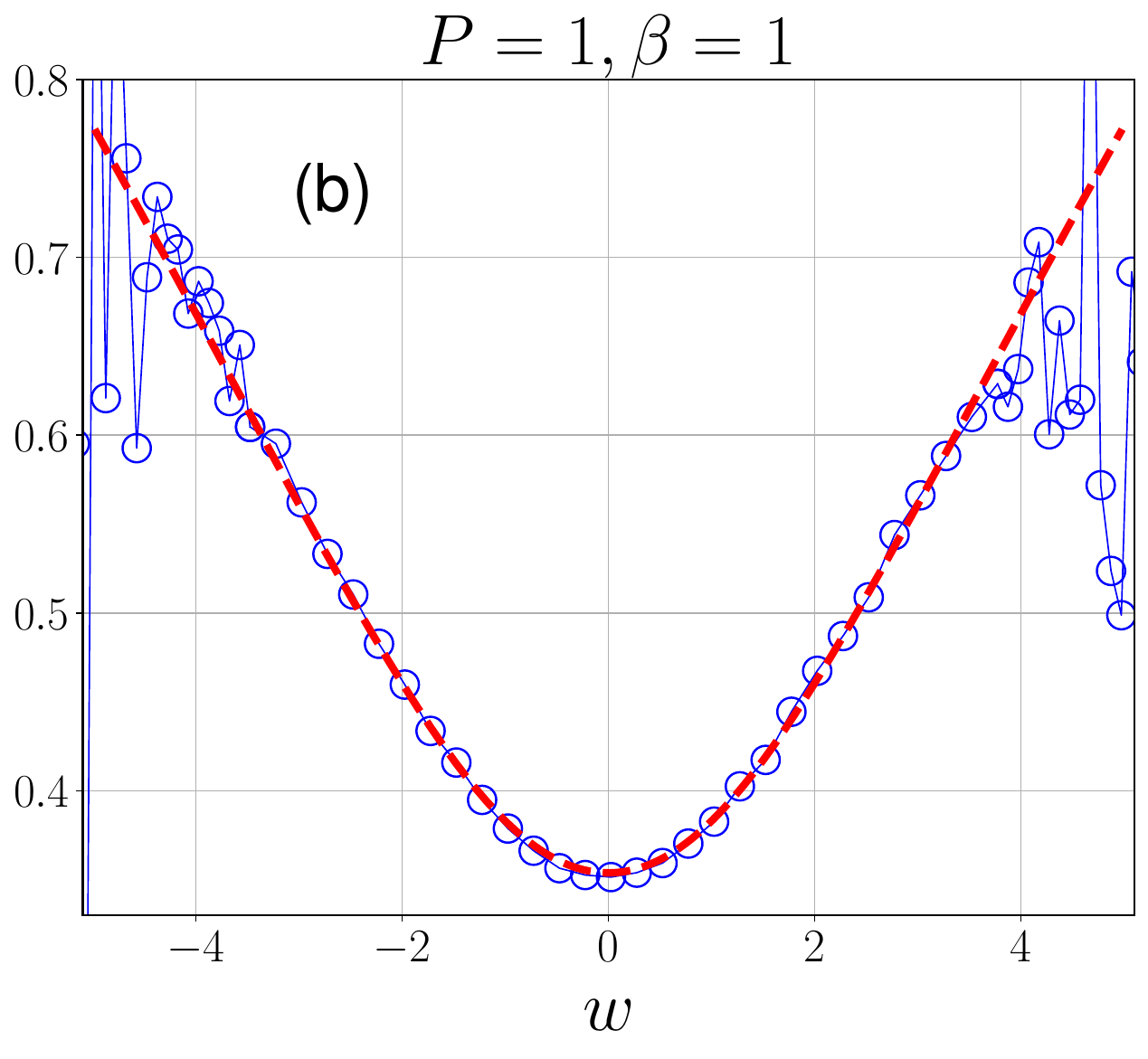}
    \includegraphics[width=4.9cm, height=5.0cm]{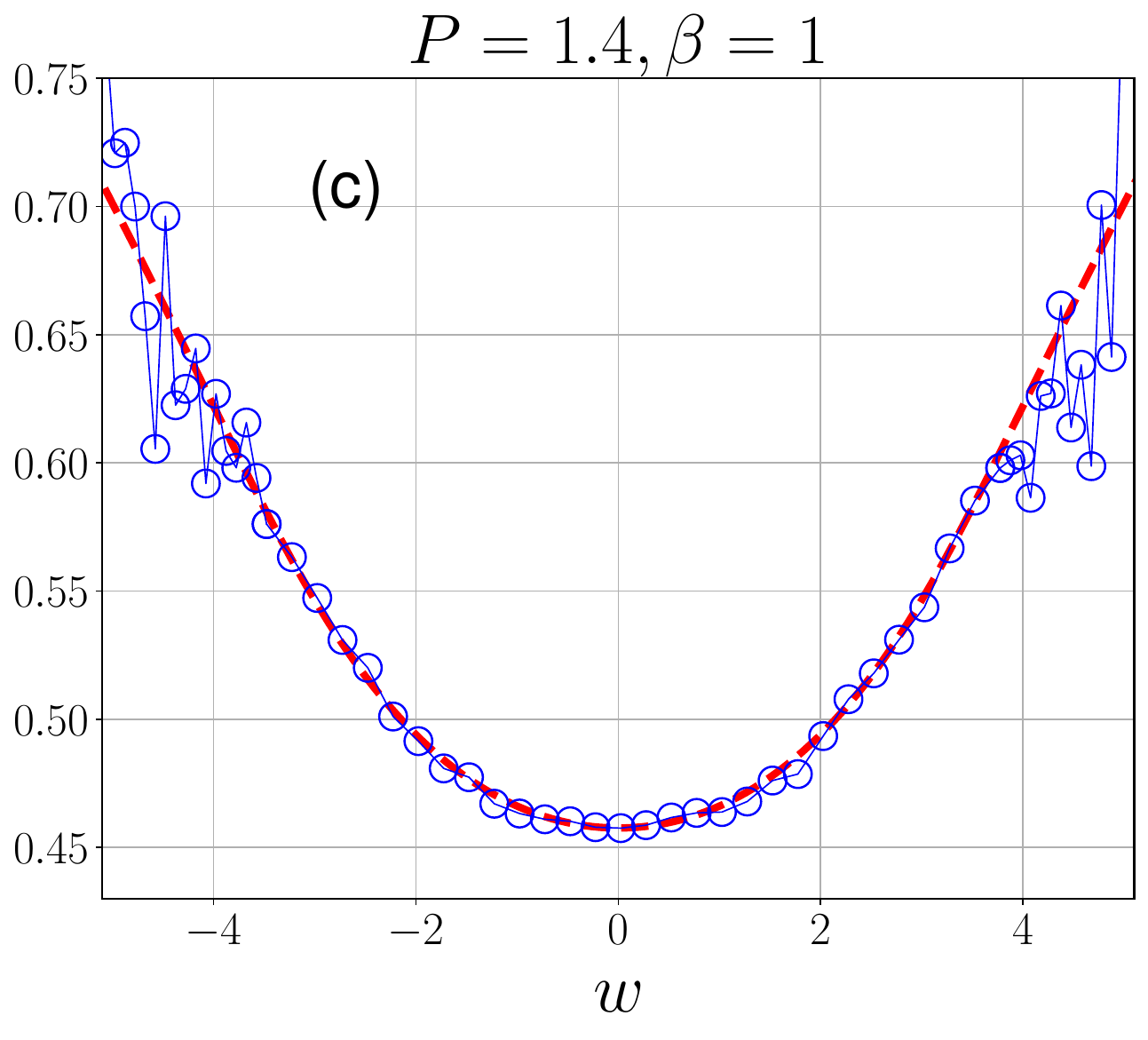}
    \caption{Plots comparing the quasiparticle diffusion coefficient $\mathfrak{D}(w)$ obtained from simulations (blue circles) with the analytical prediction of Eq.~\eqref{eq:mfrk(D)} (red dashed line). The parameter values are the same as those in Fig.~\ref{fig:DOS}.}
    \label{fig:Diffusion}
\end{figure}

\begin{SCfigure}[0.7][htbp]
    \centering
    \includegraphics[width=7.2cm, height=5.25cm]{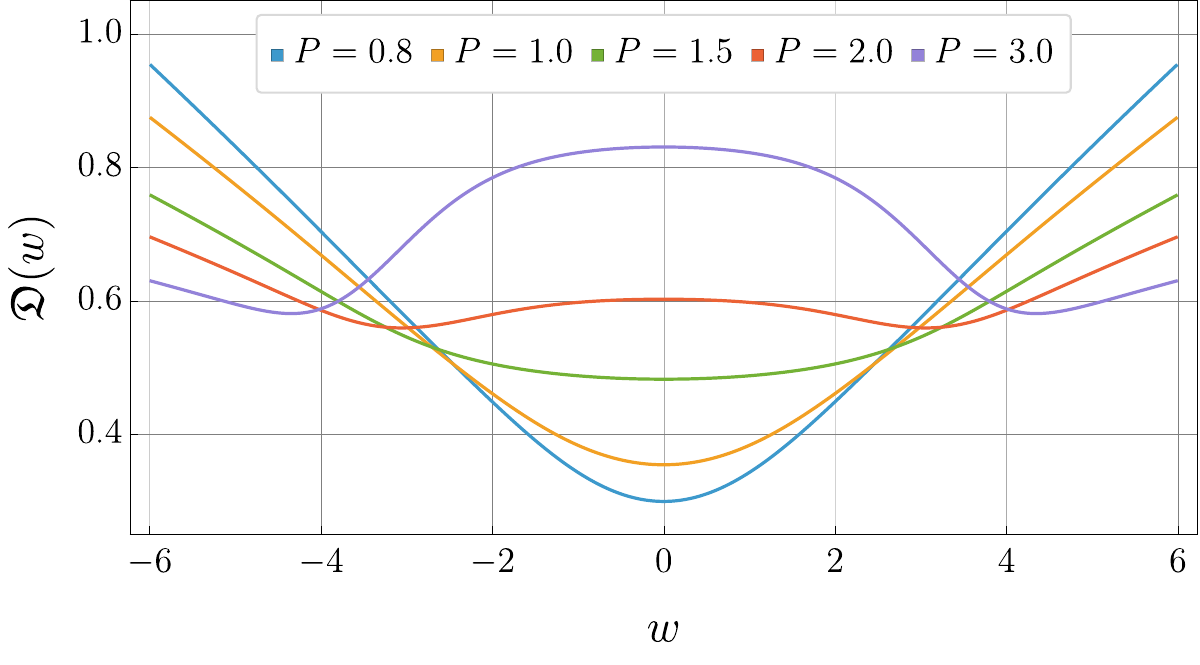}
    \caption{Quasiparticle diffusion constant $\mathfrak{D}(w)$ obtained by solving numerically Eq. \ref{eq:mfrk(D)} (see end of Sec. \ref{sec5}). The temperature is fixed $\beta=1$, and different values of pressure are represented. Note that when $P>1.46$, the average stretch is negative $\nu<0$.}
    \label{fig:tbadiff}
\end{SCfigure}

In Fig.~\ref{fig:Diffusion} we compare $\mathfrak{D}(w)$ with the hydrodynamic prediction  Eq.~\eqref{eq:mfrk(D)}, again for the three sets of parameters used before. One observes a good agreement between numerical data and theory, except for large $|w|$ where the number of binned eigenvalues is too small for achieving good statistics, see Fig.~\ref{fig:DOS}. At $w=0$ the diffusion constant 
is increasing with the pressure at fixed $\beta$, which seems physically plausible. Since simulations are time-consuming,
we use TBA for a more global picture. The respective result is plotted in Fig. \eqref{fig:tbadiff}.
Surprisingly, while the increase at $w=0$ is confirmed, the diffusion constant at large $|w|$ reverses its tendency and becomes decreasing with the pressure. 

\begin{figure}[h]
    \centering
    \includegraphics[width=4.9cm, height=5.0cm]{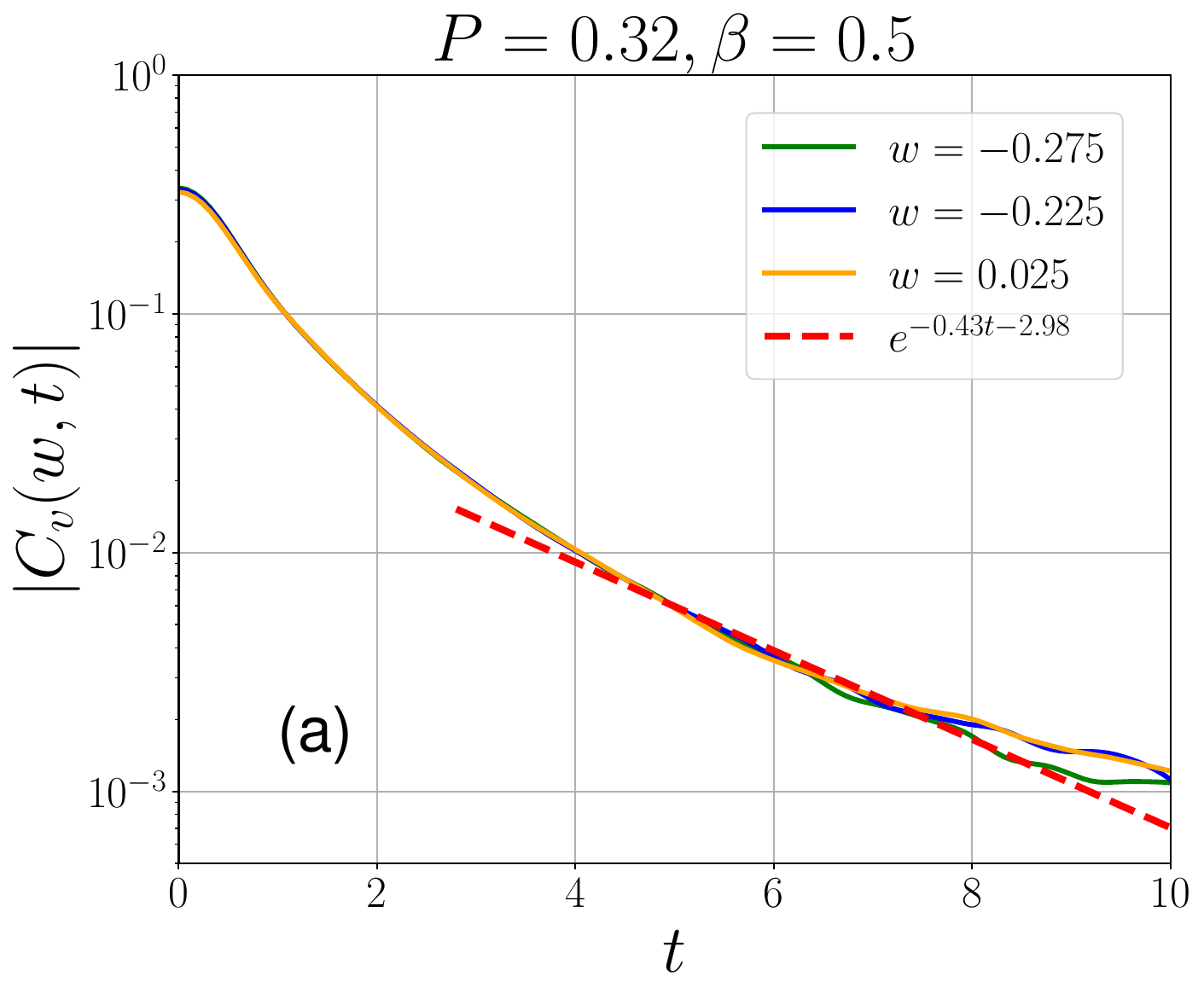}
    \includegraphics[width=4.9cm, height=5.0cm]{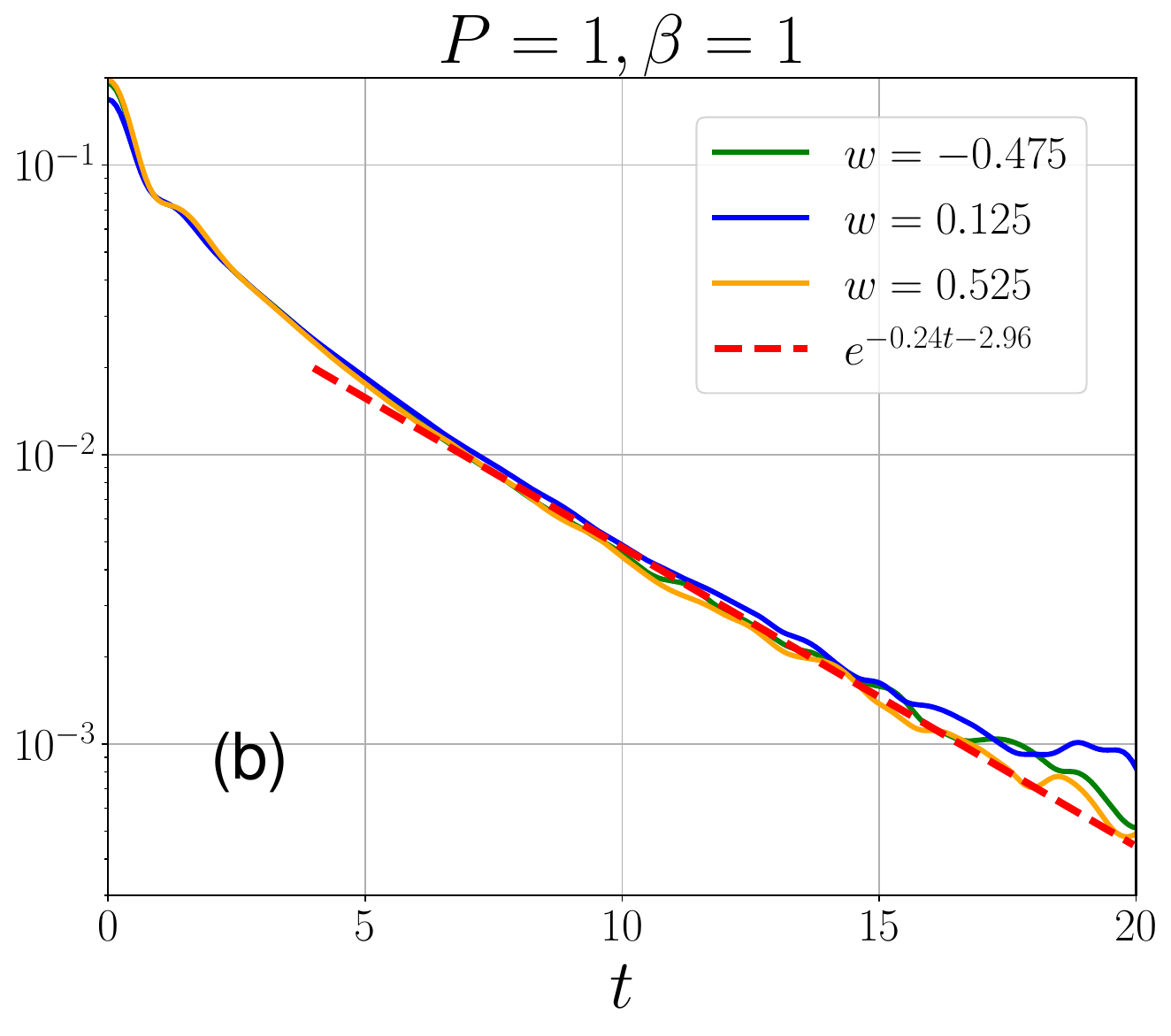}
    \includegraphics[width=4.9cm, height=5.0cm]{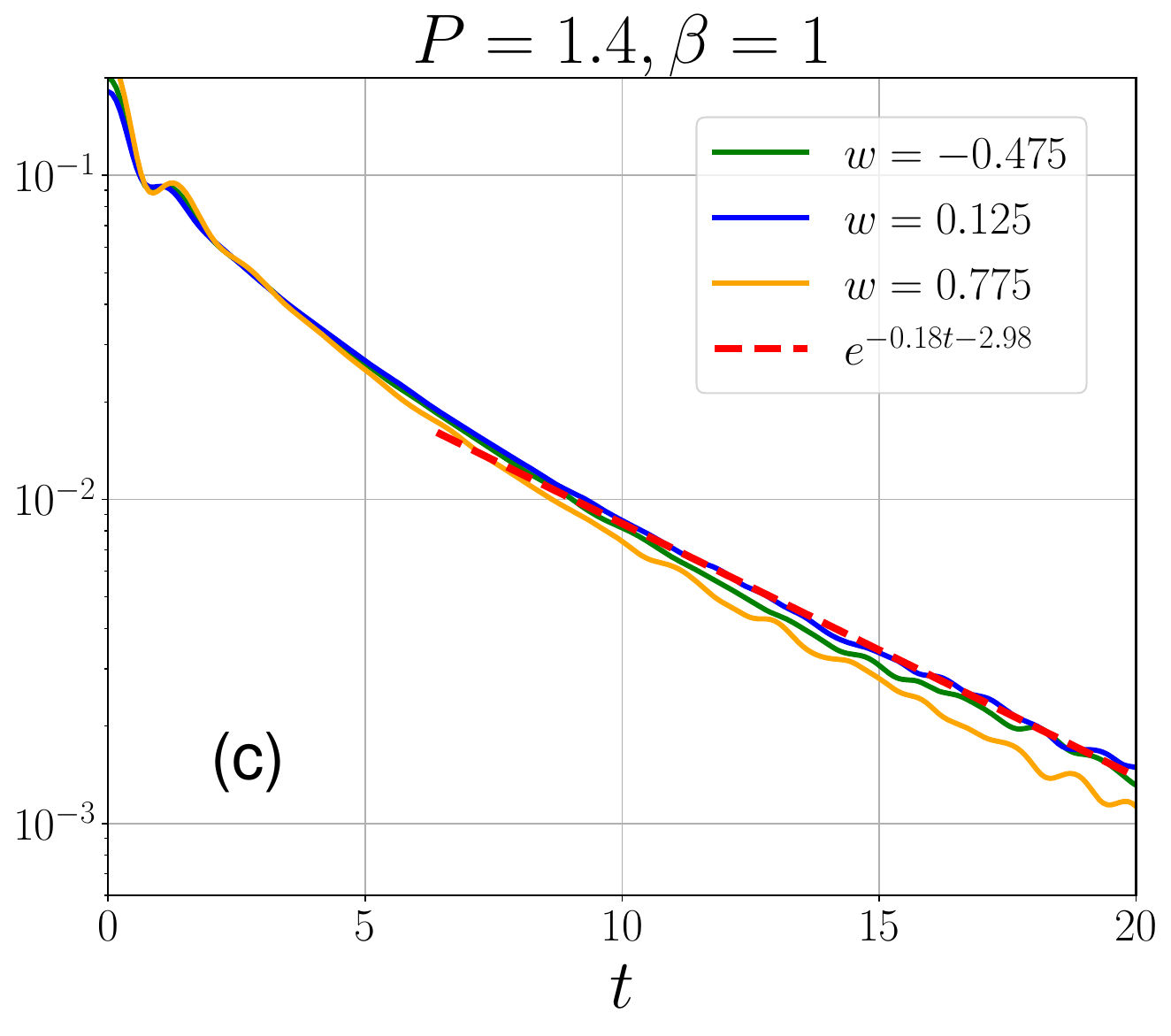}
    \caption{Plots of the quasiparticle velocity autocorrelation function, defined in Eq.~\eqref{eq:vcorr_quasi}, for different rapidities $w$ as a function of time. The simulation results (solid lines) exhibit an exponential decay and are compared with exponential fits (red dashed lines). The parameter sets are identical to those used in Fig.~\ref{fig:DOS}.}
    \label{fig:w_corr}
\end{figure}

Of great interest is the long time decay of the velocity auto-correlation function. For the Toda
fluid the decay can be deduced from Fig. \ref{fig:qpdiff}, which exemplifies how spatial fluctuations
of the initial equilibrium state are transformed to the velocity auto-correlation of the
quasiparticle. The equilibrium state of the Toda fluid is rapidly mixing, and thus one
would expect to observe an exponential decay of the time correlation. This is indeed
confirmed by the simulation results displayed in Fig. \ref{fig:w_corr}. The decay rate is decreasing for
increasing pressure.

\begin{SCfigure}[0.7][htbp]
    \centering
    \includegraphics[width=7.2cm, height=5.25cm]{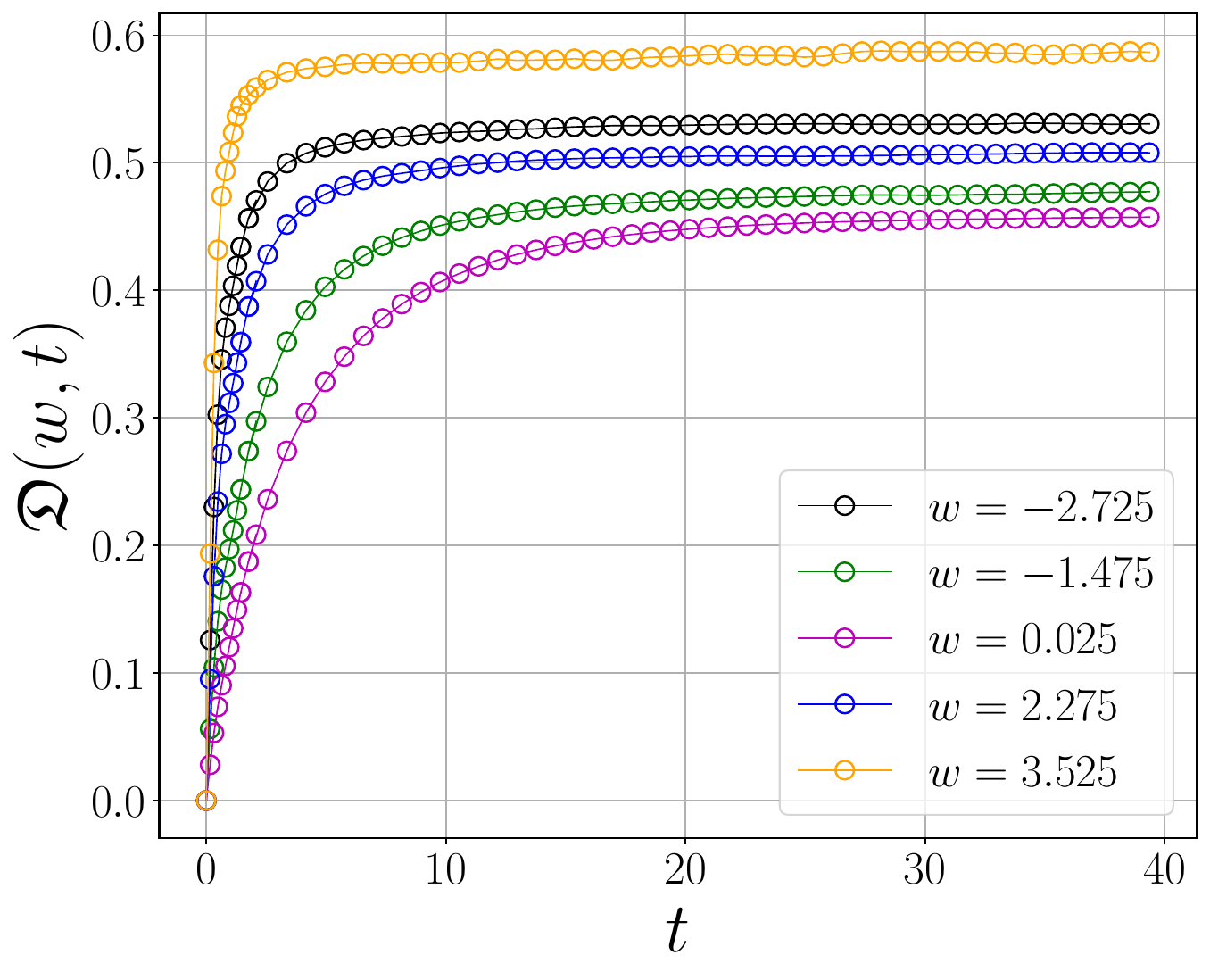}
    \caption{Plot showing the simulation result of the time-dependent diffusion constant
    $\mathfrak{D}(w,t)$ for several rapidities $w$. 
    The parameters considered are $P=1.4,~\beta=1,~N=500$, and $\mathcal{R}=10^6$.}
    \label{fig:Diff_sat}
\end{SCfigure}

In Fig.~\ref{fig:Diff_sat}, we explore in more detail the time-dependent diffusion constant $\mathfrak{D}(w,t)$ for representative  values of $w$ to study the convergence at long times. As expected, there is rapid convergence with
the exception $w=3.525$. For this rapidity the number of realizations is already at a lower margin. \\

\noindent
\textbf{Numerical solution of TBA}.
\label{sec:TBA-sol}
To evaluate the TBA predictions of $v^{\rm eff}(w)$, $\mathfrak{D}(w)$ and $D_\circ$ in Eqs.~\eqref{eq_tbaveff}, \eqref{eq:mfrk(D)} and \eqref{eq:D_circ}, respectively, we first need to solve the TBA equation~\eqref{eq_tba} for $\vartheta(w)=\exp(-\epsilon(w))$ and then use the solution in Eq.~\eqref{eq_dress} to perform the dressing operation.

Since several integral equations and integral operators are involved in the computation, we start by discretizing the rapidity space. We introduce a cutoff $\Lambda$ for the rapidity space, the validity of which must be checked \textit{a posteriori}. We discretize the interval $[-\Lambda,\Lambda]$ into $N$ equispaced points $w_j = -\Lambda + \frac{2\Lambda}{N}j$ with $j=0,1,...,N$, although non-uniform discretizations are also possible.

Next, we approximate the linear operator in Eq.~\eqref{eq_shiftop} by a matrix. To handle the divergence inside the kernel, we discretize the integral operator using the midpoint rule
\be\label{eq_Tij}
T_{ij} = 2\int_{w_j}^{w_{j+1}}dw'\log\left|\frac{w_i+w_{i+1}}{2}-w'\right| = \left[w\log|w| - w\right]\Bigg\vert_{w=\frac{w_i+w_{i+1}}{2}-w_j}^{w=\frac{w_i+w_{i+1}}{2}-w_{j+1}}.
\ee
Accordingly, functions of the rapidities $f(w)$ are also discretized at the midpoints, $f_i = f\left(\frac{w_i+w_{i+1}}{2}\right)$. The matrix-vector product $\sum_j T_{ij}f_j$ provides a good approximation of $Tf(w)$ provided that the function $f(w)$ is sufficiently smooth and can be considered constant within each cell $[w_i,w_{i+1}]$.

The TBA equation, Eq.~\eqref{eq_tba}, is thus reduced to a set of $N$ coupled algebraic equations, which can be solved via Newton iteration. While a solution can be found for any GGE, for thermal states the free energy exhibits two branches depending on the sign of $\nu$, and ensuring convergence to the $\nu<0$ branch is non-trivial. We refer to Ref.~\cite{mendl22} for further discussions. On the other hand, an exact analytical solution exists for thermal states \cite{opper85}. The filling function $\vartheta(w)$ takes the form
\be\label{eq_opper}
\vartheta_{\beta,P}(w) = \frac{\sqrt{\beta}\Gamma(P)\mathrm{e}^{-\beta w^2/2}}{\sqrt{2 \pi}P|\hat{D}_P(\sqrt{\beta}w)|^2}, \qquad
\hat{D}_P(w) = 
\int_0^\infty \mathrm{d}t \, t^{P - 1} \mathrm{e}^{\mathrm{i}wt} \mathrm{e}^{-\frac{1}{2}t^2}, 
\ee
where $\Gamma(P)$ is the Euler Gamma function evaluated at the given value of the pressure. Once the filling function is known, all relevant thermodynamic quantities can be readily obtained via the dressing procedure. Note that  the dressing operator in Eq.~\eqref{eq_dress} is now simply the matrix multiplication $f^{\rm dr}(w) = (1-T\vartheta)^{-1}f(w)$. Using Eq.~\eqref{eq_tbaveff} and Eq.~\eqref{eq:mfrk(D)} one can compute the effective velocity and self-diffusion constant, respectively. A working Mathematica code for evaluating the TBA solutions, along with python code for molecular dynamics simulations, is provided in a GitHub repository \cite{chahal2026toda}.


\section{Diffusion of eigenvectors}
\label{sec6}
\textbf{Theory predictions}. An eigenvector is tagged by its time-independent eigenvalue. As illustrated already in Figure 2, the eigenvector shows an intrinsic blurriness and propagates with a definite velocity superimposed by small random fluctuations.
In this section we study such features in more detail. The underlying evolution equations are stated in Eqs. \eqref{eq_eomF},
which are viewed as a chain with local variables $(p_j, a_j)$. There is still a hydrodynamic description. However,  the effective
velocity is modified to  
\begin{equation}\label{8.1}
 \tilde{v}^\mathrm{eff}(w) = \nu^{-1} \big(v^\mathrm{eff}(w) - \mathsfit{q}\big), \quad \mathsfit{q} = \nu\int_\mathbb{R}  dw \rho_\mathsf{p}(w)w.
\end{equation}
To emphasize the respective correspondence, the same quantity with no tilde refers to quasiparticles, while with tilde means eigenvectors.  $\tilde{v}^\mathrm{eff}$ can also be written as the solution of 
\begin{equation}\label{6.18}
\tilde{v}^\mathrm{eff}(w) 
 = \nu^{-1}(w - \mathsfit{q}) +2 \nu^{-1}\int_\mathbb{R} \mathrm{d}w' \rho_\mathrm{DOS}(w')\log|w-w'| \big(\tilde{v}^\mathrm{eff}(w') - \tilde{v}^\mathrm{eff}(w)\big). 
\end{equation} 
Thus only the bare velocity is modified. 
Note that the subtraction ensures that
\begin{equation}\label{6.18}
\int_\mathbb{R}  dw \rho_\mathsf{p}(w)\tilde{v}^\mathrm{eff}(w) = 0.
\end{equation}
In \cite{spohn24}, Section $5.2$, this modification is discussed for the chain version of hard rods. The Toda lattice is covered in 
 \cite{spohn24}, Section 6.3, using TBA identities.

By analogy, one would expect that the mean velocity of an eigenvector equals $ \tilde{v}^\mathrm{eff}$. To confirm, we  start from the chain $[1,\dots,N]$ with periodic boundary conditions. In thermal equilibrium the eigenvectors of the closed Lax matrix are localized mostly within a few lattice sites.  The location of the normalized eigenvector is defined by the first moment of  $|\psi_\alpha(j,t)|^2$ as
\begin{equation}\label{8.2} 
\tilde{Q}_\alpha(t) = \sum_{j=1}^Nj |\psi_\alpha(j,t)|^2.
\end{equation}
Strictly speaking, since $j$  has a jump of size $N$ across the bond $(N,1)$, this formula makes sense only for $\tilde{Q}_\alpha$ a few lattice sites away from either $1$ or $N$. Upon taking the time derivative, one obtains the commutator $[B_N,\mathrm{diag}(\{j\})]$ and hence the eigenvector velocity,
\begin{equation}\label{8.3} 
\frac{\mathrm{d}}{\mathrm{d}t} \tilde{Q}_\alpha(t) = \langle  \psi_\alpha(t), \tfrac{1}{2}((L_N)^\uparrow + (L_N)^\downarrow)\psi_\alpha(t)\rangle = \tilde{v}_\alpha(t).
\end{equation}
Here $(L_N)^\uparrow$ is the strictly upper triangle and $(L_N)^\downarrow$ the strictly lower triangle of $L_N$.  We regard \eqref{8.3} as definition of the velocity for all $\alpha = 1,\dots,N$. The eigenvector location is then the time integral which yields the motion on the circle,  including its proper winding. 

To determine the GGE averaged velocity of an eigenvector is even simpler than the corresponding argument for a quasiparticle.
We observe the identities
\begin{eqnarray}\label{8.4}
 &&\frac{1}{N} \sum_{\alpha=1}^N \langle (\lambda_\alpha)^n \tilde{\mathsfit{v}}(\lambda_\alpha)\rangle_{P,V}
   =  \frac{1}{N} \sum_{\alpha=1}^N \big\langle(\lambda_\alpha)^n   \langle\psi_\alpha|  \tfrac{1}{2}((L_N)^\uparrow + (L_N)^\downarrow)\psi_\alpha\rangle\big\rangle_{P,V},\nonumber\\
 && =\frac{1}{N} \langle \mathrm{tr}[(L_N)^n  \tfrac{1}{2}((L_N)^\uparrow+ (L_N)^\downarrow)] \rangle_{P,V}.
 \end{eqnarray}
 Taking the  large $N$ limit on either sides results in 
\begin{equation}
\int_\mathbb{R} \mathrm{d} \lambda \rho_\mathrm{DOS}(\lambda) \tilde{\mathsfit{v}}(\lambda)  \lambda^n,
=  \int_\mathbb{R} \mathrm{d} \lambda \rho_\mathrm{p}(\lambda)(v^\mathrm{eff}(\lambda) - \mathsfit{q}) \lambda^n,
\end{equation}
using Eq. (6.18) in \cite{spohn24}. In conclusion
\begin{equation}\label{8.5} 
 \nu^{-1} (v^\mathrm{eff}(\lambda) - \mathsfit{q}) =\tilde{\mathsfit{v}}(\lambda),
\end{equation} 

The computation of the diffusion constant of eigenvectors is more difficult and does not seem to be available in the current literature. 
Kindly enough Amol Aggarwal  pointed out to us that this quantity can be deduced from his limit theorems \cite{aggarwal26} with the result
\begin{equation}\label{8.6} 
\tilde{\mathfrak{D}}(w) = \frac{1 }{2\nu^3\rho_\mathsf{s}(w)^{2}}\int~dw' \rho_\mathrm{DOS}(w')\big(\nu \rho_\mathsf{s}(w)\rho_\mathsf{s}(w') -\varphi^\mathrm{dr}(w,w') \big)^2|v^\mathrm{eff}(w)-v^\mathrm{eff}(w')|.
\end{equation} 

Comparing with the quasiparticle diffusion constant Eq.~\eqref{eq:mfrk(D)}, one notes that the quadratic variation of $v^\mathrm{eff}$ appears
in both equations except for the prefactor $\nu^{-3}$ which reflects the conversion between the density of eigenvalues and quasiparticles. 
 In addition, Eq. \eqref{8.6} contains the quadratic variation of $\mathsfit{q}$ and the respective cross term.\\

\noindent
\textbf{Numerical results}. For numerical evaluation we closely follow the same procedure  as used for quasiparticles  in Sec. \ref{sec5}. We generate the initial conditions according to the thermal equilibrium ensemble and evolve the Toda particles in time using  molecular dynamics.
For each configuration of the particles, the associated Lax matrix is diagonalized to obtain its eigenvalues and the corresponding eigenvectors. The eigenvectors are then evolved using Eq.~\eqref{eq_lin}. The instantaneous velocity associated with a given eigenvector is then computed from Eq.~\eqref{8.3}, in analogy to quasiparticle velocity obtained from Eq.~\eqref{eq_qpvel}. Finally, since eigenvalues vary from one realization to another, the same spectral binning and averaging procedure described in Sec. \ref{sec5} is employed to obtain spectrally resolved results for the eigenvector dynamics and its diffusion constant. More precisely, to compute the diffusion constant we first compute the truncated velocity autocorrelation of eigenvectors, similar to what is done for quasiparticles in Eq.~\eqref{eq:vcorr_quasi}. Finally, integrating velocity autocorrelation over time (as done for quasiparticles in Eq.~\eqref{D(w,t)-qp}) yields the eigenvector diffusion constant. 

 \begin{figure}[t]
    \centering
      \includegraphics[width=7.2cm, height=5.25cm]{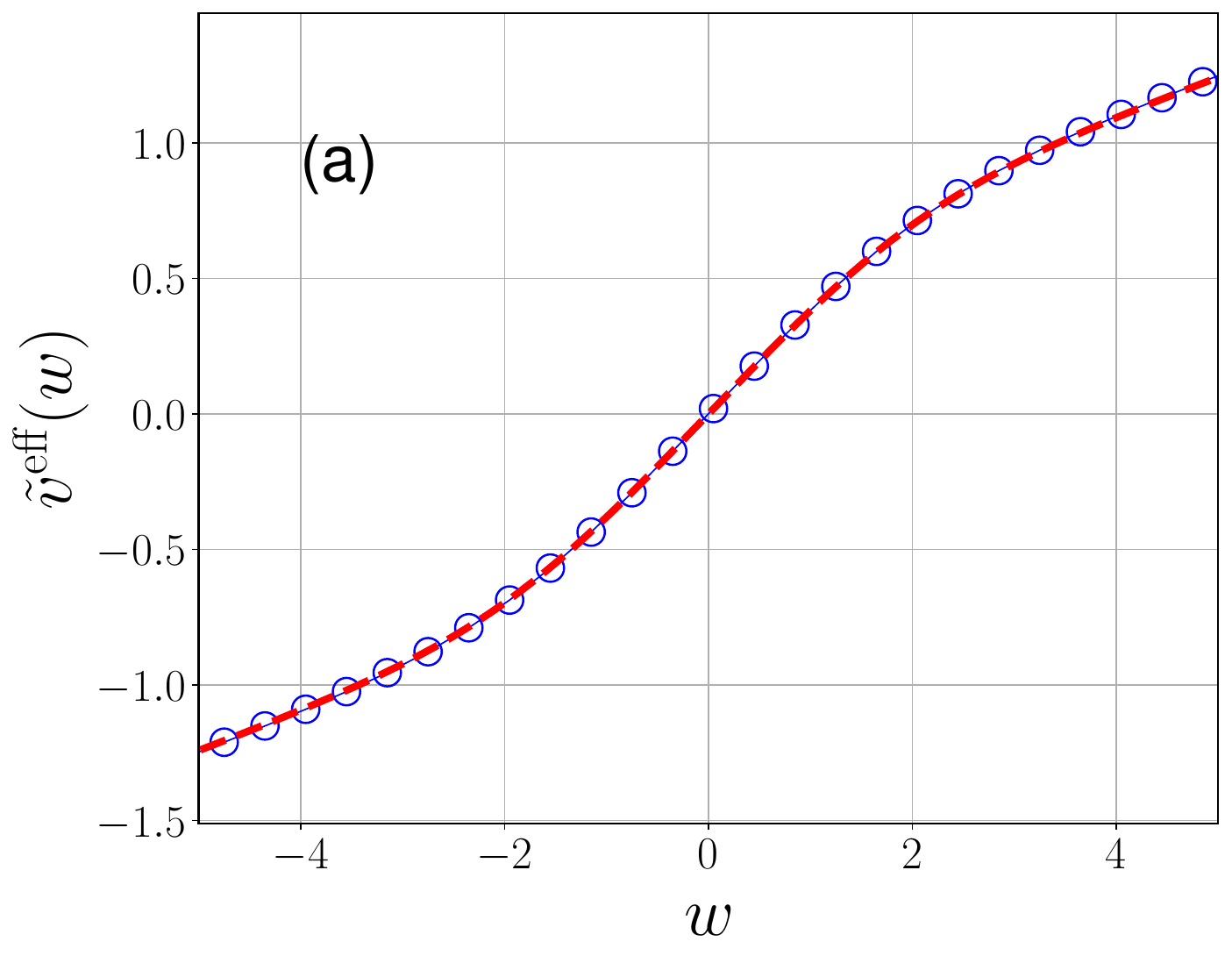}
      \includegraphics[width=7.2cm, height=5.25cm]{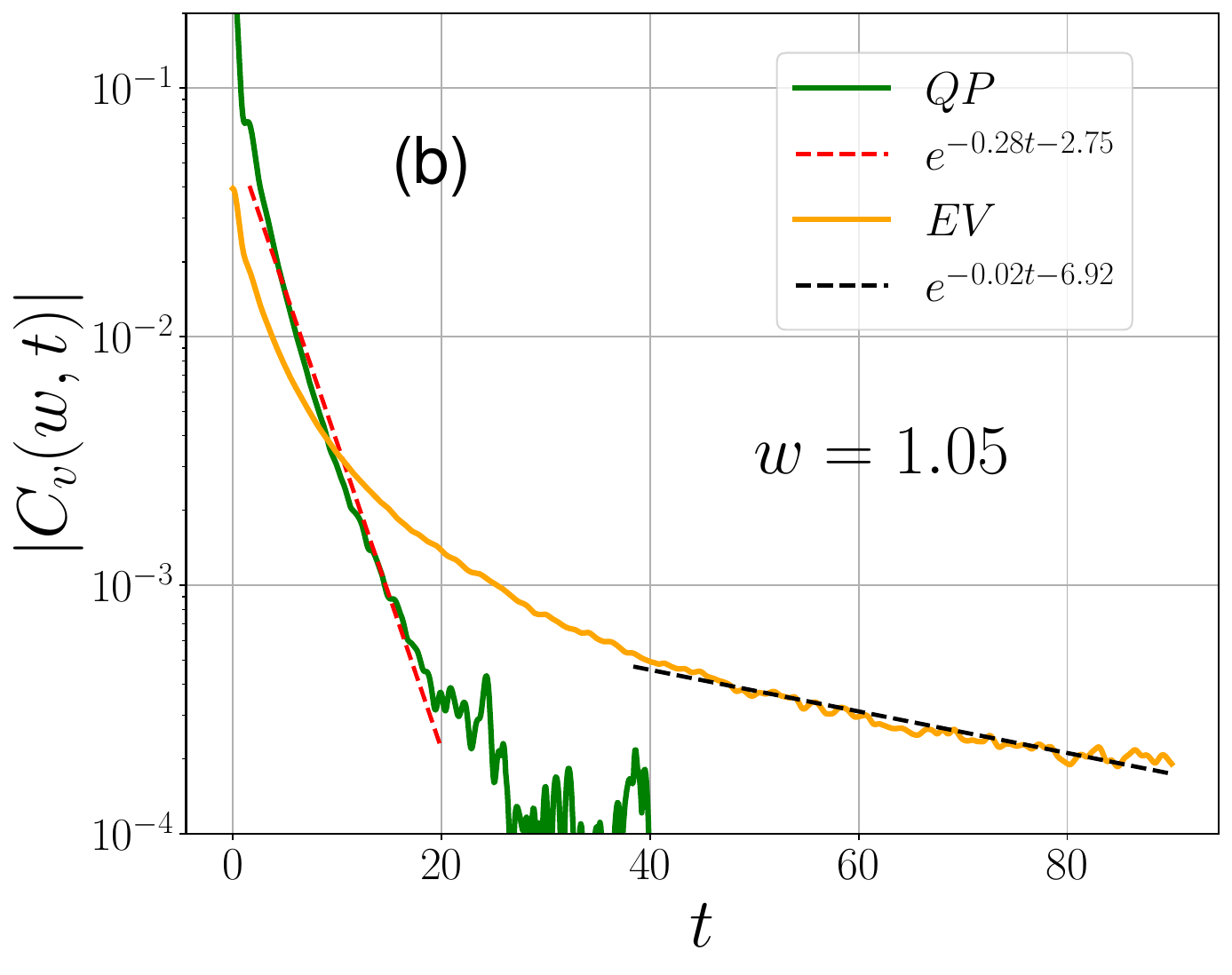}
    \includegraphics[width=7.2cm, height=5.25cm]{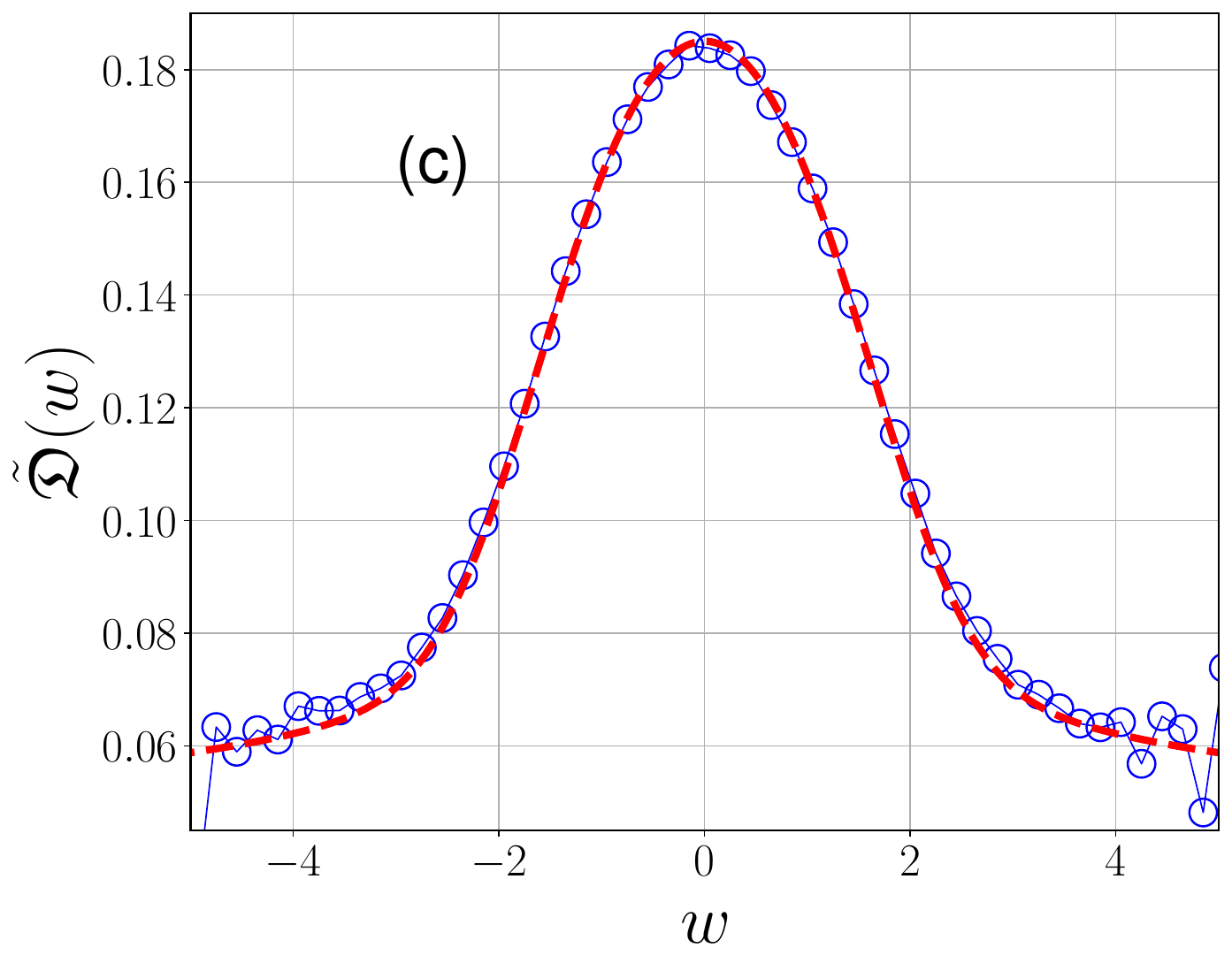}
    \includegraphics[width=7.2cm, height=5.25cm]{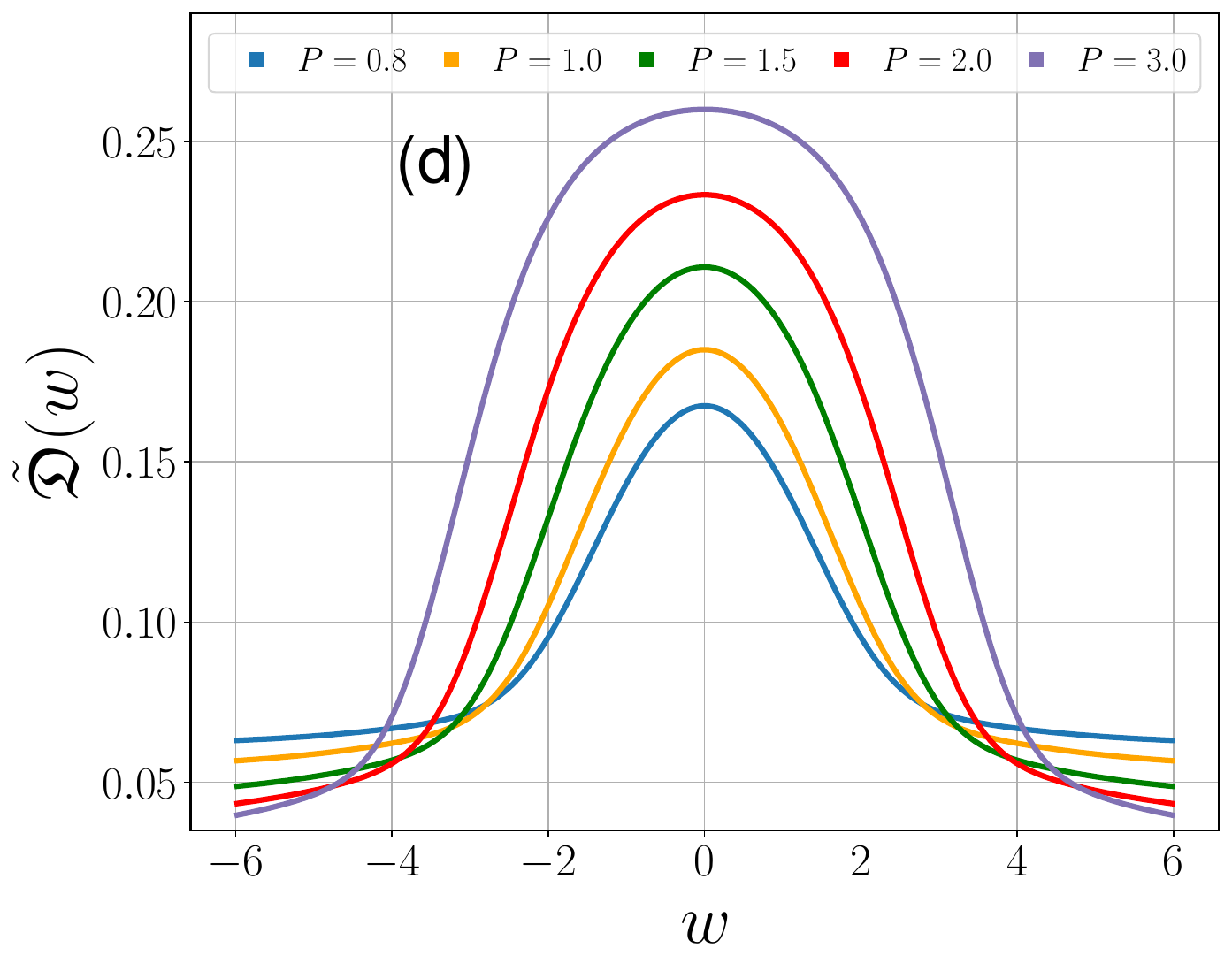}
    \caption{Comparison between numerical and analytical results for eigenvector dynamics. Panel (a) shows the effective velocity of eigenvectors (blue circles) and  compares with the analytical result Eq.~\eqref{8.5} (red dashed line).  For given eigenvalue, panel (b) compares the decay of velocity auto-correlation functions for  quasiparticle (QP) and eigenvector (EV). Their exponential decay is demonstrated by the respective exponential fits, shown as dashed lines. Panel (c) compares eigenvector diffusion constant (blue circles) with the analytical prediction Eq.~\eqref{8.6} (red dashed line). The parameters for (a), (b) and (c) are $P=1$, $\beta=1$, $N=500$, and $\mathcal{R}=10^6$. The numerically evaluated analytical expression in Eq. \eqref{8.6} is displayed in panel (d), where the eigenvector diffusion constant $\tilde{\mathfrak{D}}(w)$ {\emph vs.} $w$ is plotted for $\beta=1$ and several values of $P$. Note that for $P>1.46$, the average stretch is negative, $\nu<0$.}
    \label{fig:EV_diff}
\end{figure}

The results of these numerical calculations are presented in Fig.~\ref{fig:EV_diff} in four panels. In panel (a) we plot the effective velocity $\tilde v^{\rm eff}(w)$ from Eq.~\eqref{8.1} along with numerical data (circles), where we observe excellent agreement. Note, for our thermal initial state $\mathsfit{q}=0$. In panel (b) we show how the velocity autocorrelation of an eigenvector (orange line) decays with time. For comparison, we also plot quasiparticle velocity autocorrelation for the same $w=1.05$. We observe, that the eigenvector velocity auto-correlation also decays exponentially however  more slowly than the corresponding quasiparticle velocity auto-correlation. Integrating such correlations numerically over time we obtain the diffusion constant $\tilde{\mathfrak{D}}(w)$ which we compare with analytical prediction in Eq.~\eqref{8.6}. Very surprisingly, the $w$ dependence of $\tilde{\mathfrak{D}}(w)$ is upside down to that of $\mathfrak{D}(w)$ in Fig.~\ref{fig:Diffusion}, namely the curves are bent oppositely.  To have a more global view, we solve numerically the 
the analytical expression Eq.~\eqref{8.6} with $\beta=1$ and  different values of $P$. Panel (d) should be compared with Fig.~\ref{fig:tbadiff}). Apparently the diffusion constants depend smoothly on $P,\beta$. but quasiparticle and eigenvector are distinctly different. 


\section{Tagged particle diffusion}\label{sec7}
\textbf{Theoretical results}. For a system of hard rods the diffusion of a tagged particle has been studied a long time ago \cite{lebowitz68} and an exact formula for the thermal velocity auto-correlation function was obtained. Its time integral yields a simple explicit formula for the diffusion constant. But in addition, it was discovered that the velocity auto-correlation function decays slowly as $t^{-3}$. As a general experience, on a macroscopic scale integrable fluids have the same behavior. 
Thus our interest is to find out whether for the Toda fluid a corresponding result would be valid,
thereby enlarging the range of observables handled by GHD. In brackets we note that the tagged particle motion has also been studied numerically for non-integrable one-dimensional fluids \cite{Sanjib_Abhisek_tagged2013,Sanjib_Abhisek_tagged2015}. The velocity auto-correlation function shows power law decay, but the exponent seems to depend on the particular system, including examples with non-integrable decay. 

 \begin{figure}[t]
    \centering
      \includegraphics[width=4.9cm, height=5.0cm]{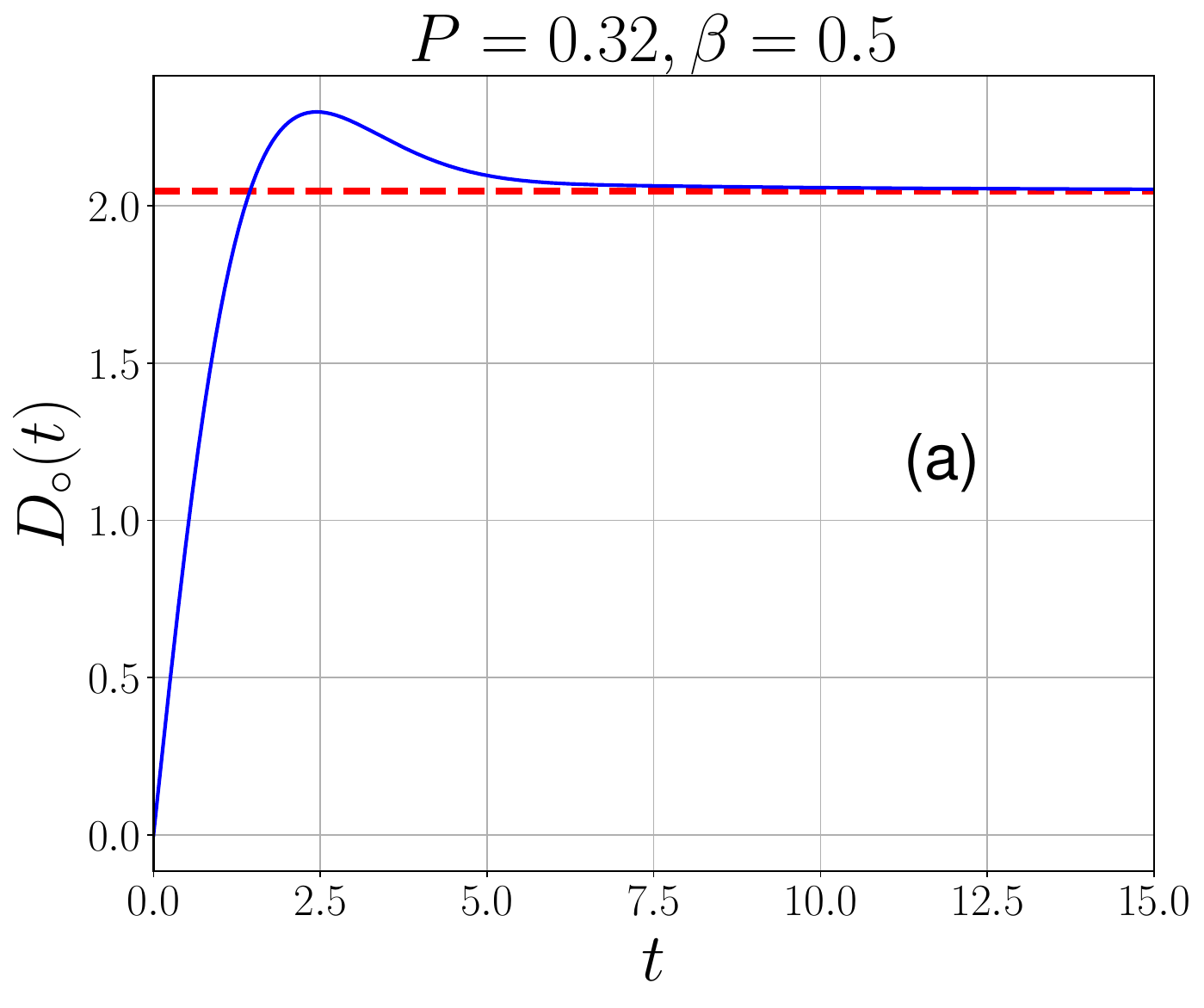}
      \includegraphics[width=4.9cm, height=5.0cm]{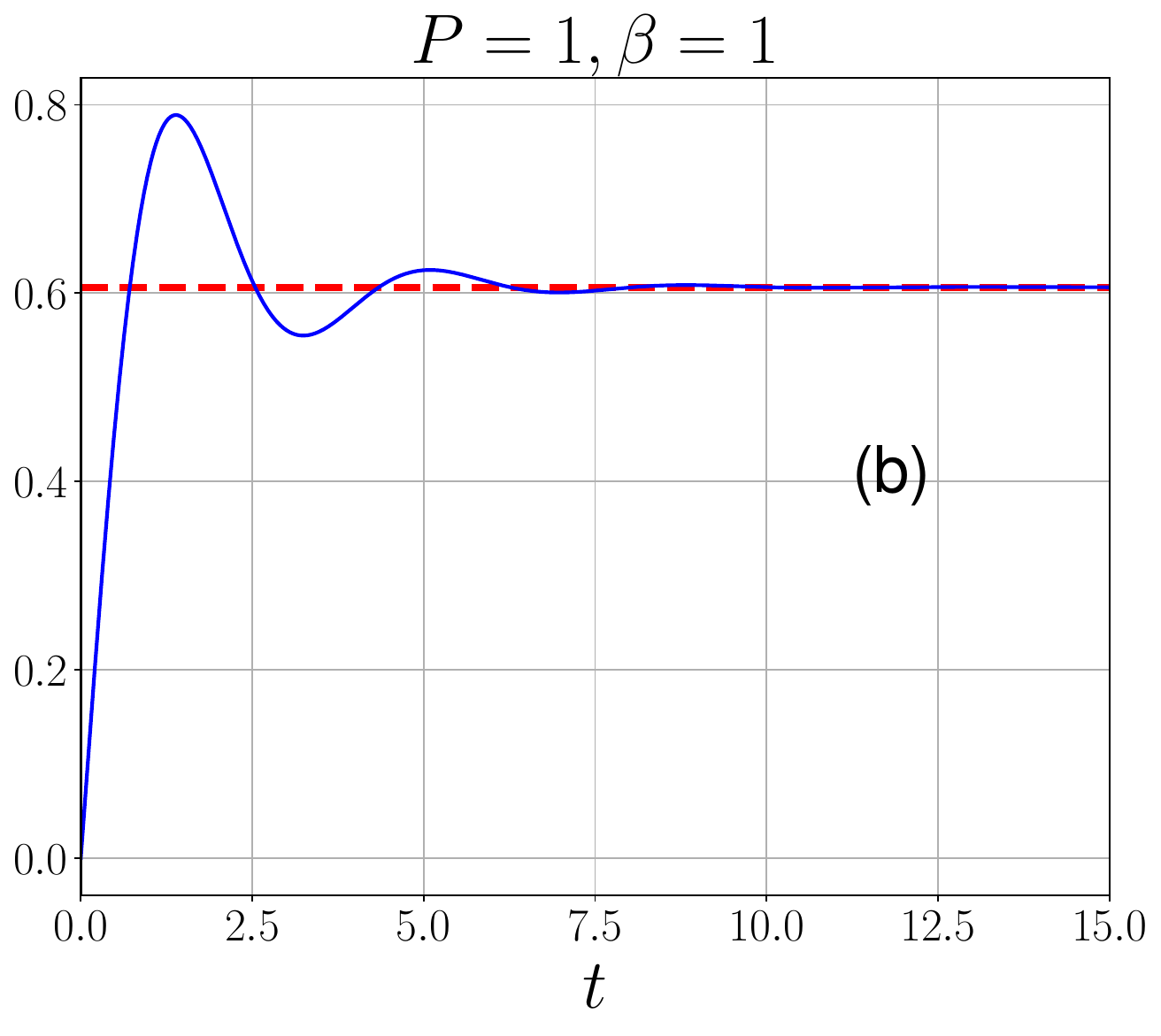}
    \includegraphics[width=4.9cm, height=5.0cm]{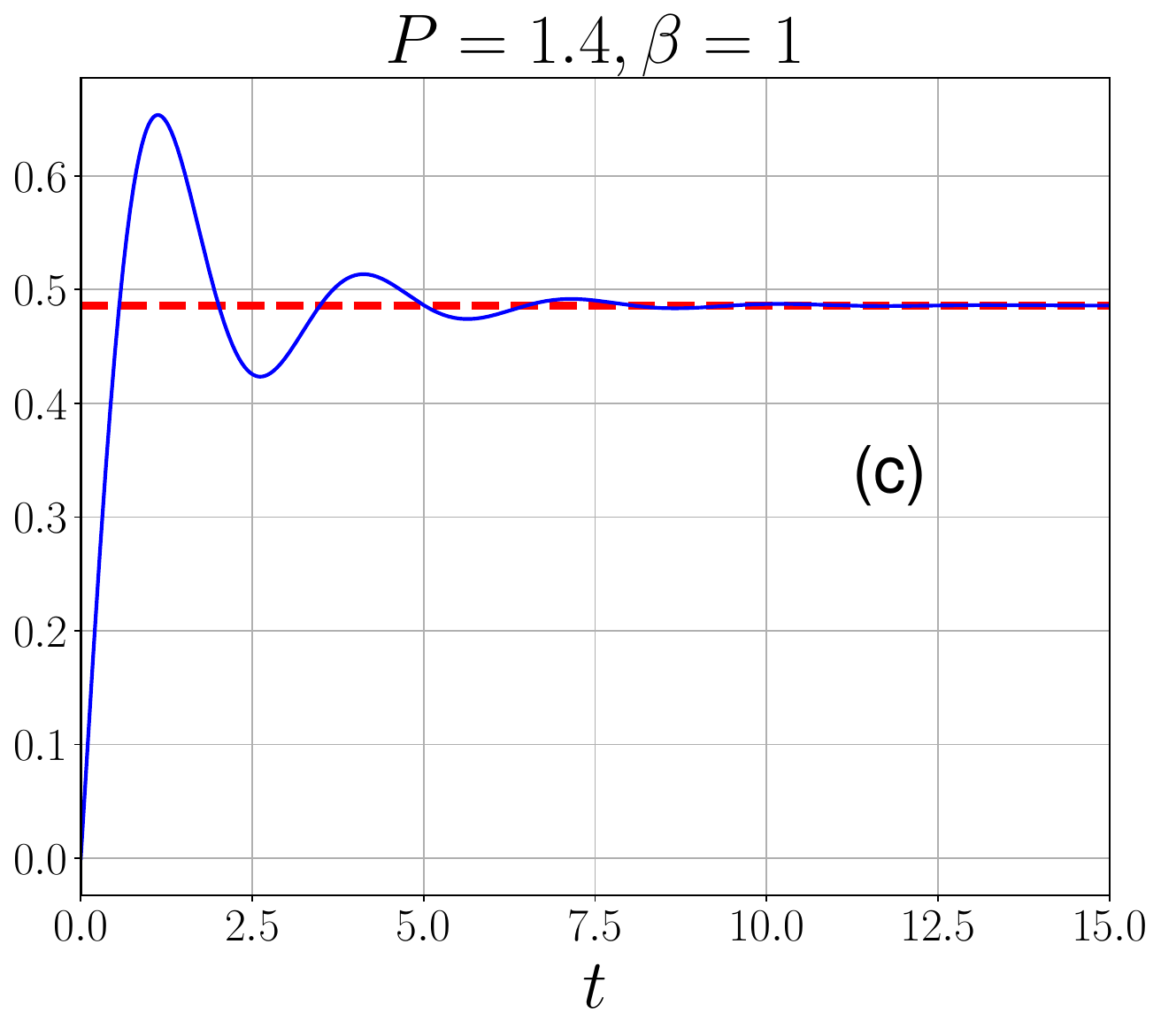}
    \caption{Plots showing the simulation results (blue line) for the time-dependent diffusion coefficient of a tagged particle and its convergence to the analytical prediction (red dashed line) given by Eq.~\eqref{eq:D_circ}. The parameter values $P$ and $\beta$ are the same as those in Fig.~\ref{fig:DOS}, while $N=10^3$ and $\mathcal{R} = 10^6$.}
    \label{fig:p_diff}
\end{figure}

The tagging of a particle poses no problem.
Hence we follow the particle $q_0(t)$ and assume $q_0(0)=0$. Without loss of generality, one can set $\langle p_0(0)\rangle_{P,V} = 0$, which remains so for all $t$.  The particle diffusion constant is then given by
\be \label{eq:diff_const_particle}
D_\circ =  \int_\mathbb{R} dt~\langle p_0(t)p_0(0)\rangle_{P,V}.
\ee
The particle momentum $p_0(t)$ is the density of total momentum field at $j=0$.  Its entire spacetime correlator  
is given by $S_j(t) = \langle p_j(t)p_0(0)\rangle_{P,V}$. As discussed in \cite{mazzuca23}, under ballistic scaling this correlator converges to 
\be \label{eq:D_o}
S(x,t) = \int_\mathbb{R} dw \delta(x - tv^\mathrm{eff}(w))  \rho_\mathsf{p}(w)([w]^\mathrm{dr}(w))^2.
\ee
We point out that depending on whether the Toda system is viewed as chain or a fluid, there are two natural choices for the grand canonical ensemble (see \cite{Doyon19}). Ref. \cite{mazzuca23} presents results based on the chain interpretation, while here we are using the fluid one. 

For the integral over time one finds 
\be \label{eq:D_circ}
D_\circ =  \int_\mathbb{R} dt S(x,t) = \int_\mathbb{R} dw \,\rho_\mathsf{p}(w)|v^\mathrm{eff}(w)|(\rho_\mathsf{s}(w))^2,
\ee
using that $v^\mathrm{eff} = [w]^\mathrm{dr}/[1]^\mathrm{dr}$. The fact that there is no dependence on $x$ reflects the rigidity of the particle motion. The choice $x=0$ would approximately single out $q_0(t)$. Equation \eqref{eq:D_circ} agrees with a very recently proved theorem, see  Eq. (2.22) of \cite{aggarwal26}. 
 
Our argument can be pushed to obtain qualitatively the slow decay $t^{-3}$. Close to $w=0$,
in good approximation, the effective velocity is linear, 
$v^\mathrm{eff}(w) \simeq m_\circ w$. Inserting in \eqref{eq:D_o} one obtains
\be
S(x,t) \simeq (1/m_\circ t) f(x/m_\circ t),
\ee
where $f(w) =  \rho_\mathsf{p}(w)[w]^\mathrm{dr}(w)^2$. By definition $f(0) = 0$ and 
$f(w) = f(-w)$.  The numerical plots in \cite{mazzuca23}, confirm that  $f(w) = c_\circ w^2$, $c_\circ >0$, for small $w$. In fact at larger pressure the curvature is very flat. Hence to this order
\be
S(x,t) \simeq  \frac{c_\circ x^2}{(m_\circ t)^3},
\ee
which signals the $t^{-3}$ power law. Our argument is less stringent than the one for the diffusion constant. Only through numerical simulations one learns more.\\\\
\textbf{Molecular dynamics}.
 To obtain the tagged particle diffusion coefficient $D_\circ$, we follow the same route as for a quasiparticle. The velocity auto-correlation is now the thermal average of  $p_0(t)p_0(0)$. No binning and truncation  are required. For a system of size $N$, the average is over $\mathcal{R}$ independent initial realizations. In  Fig.~\ref{fig:p_diff} we display the time-dependent diffusion constant $D_\circ(t)$ for the same thermodynamic parameters as used before. One observes a good convergence to $D_\circ(\infty)$ and this value agrees very well with the theory.

\begin{figure}
    \centering
    \includegraphics[width=4.9cm, height=4.8cm]{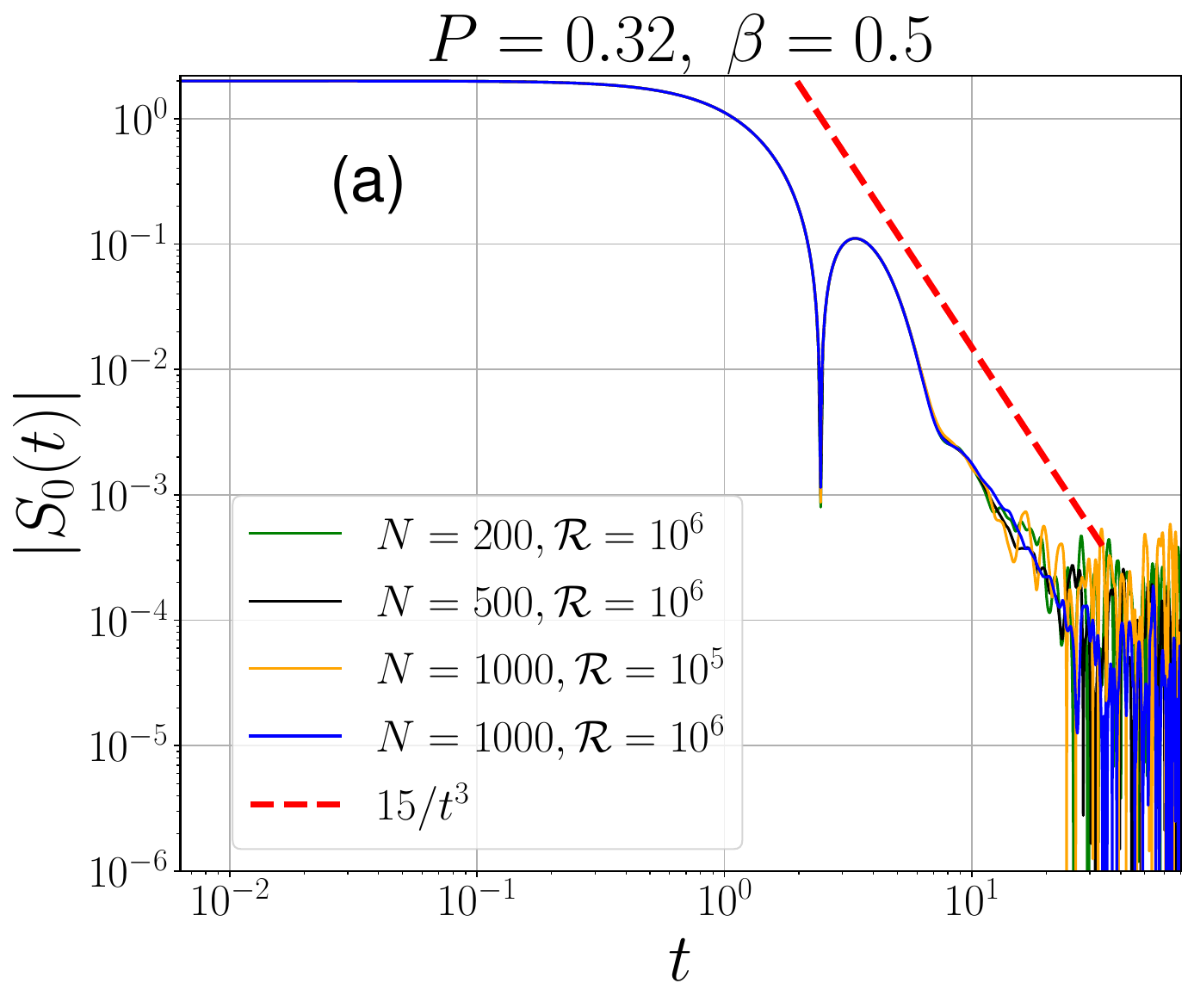}
    \includegraphics[width=4.9cm, height=4.8cm]{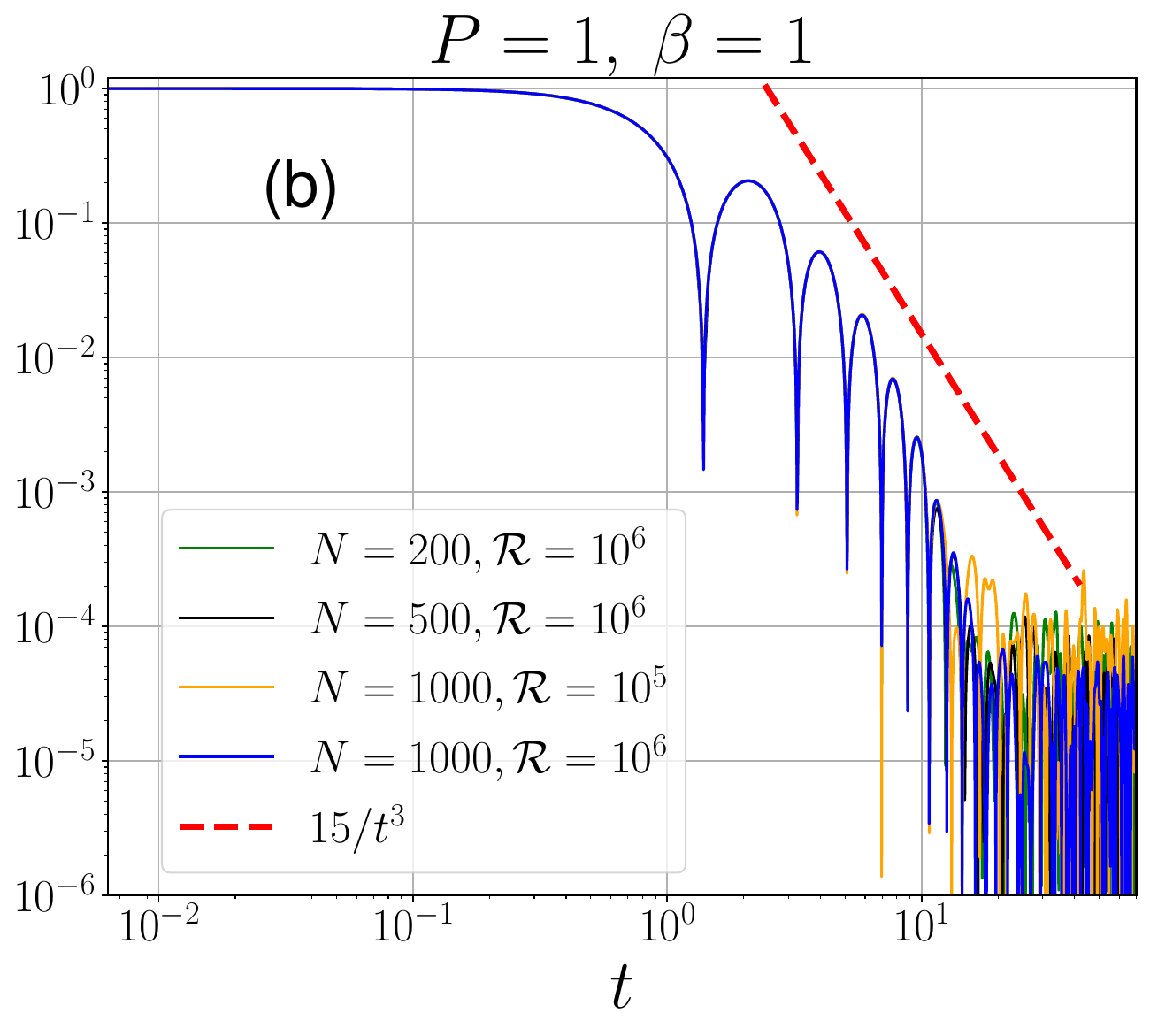} 
    \includegraphics[width=4.9cm, height=4.8cm]{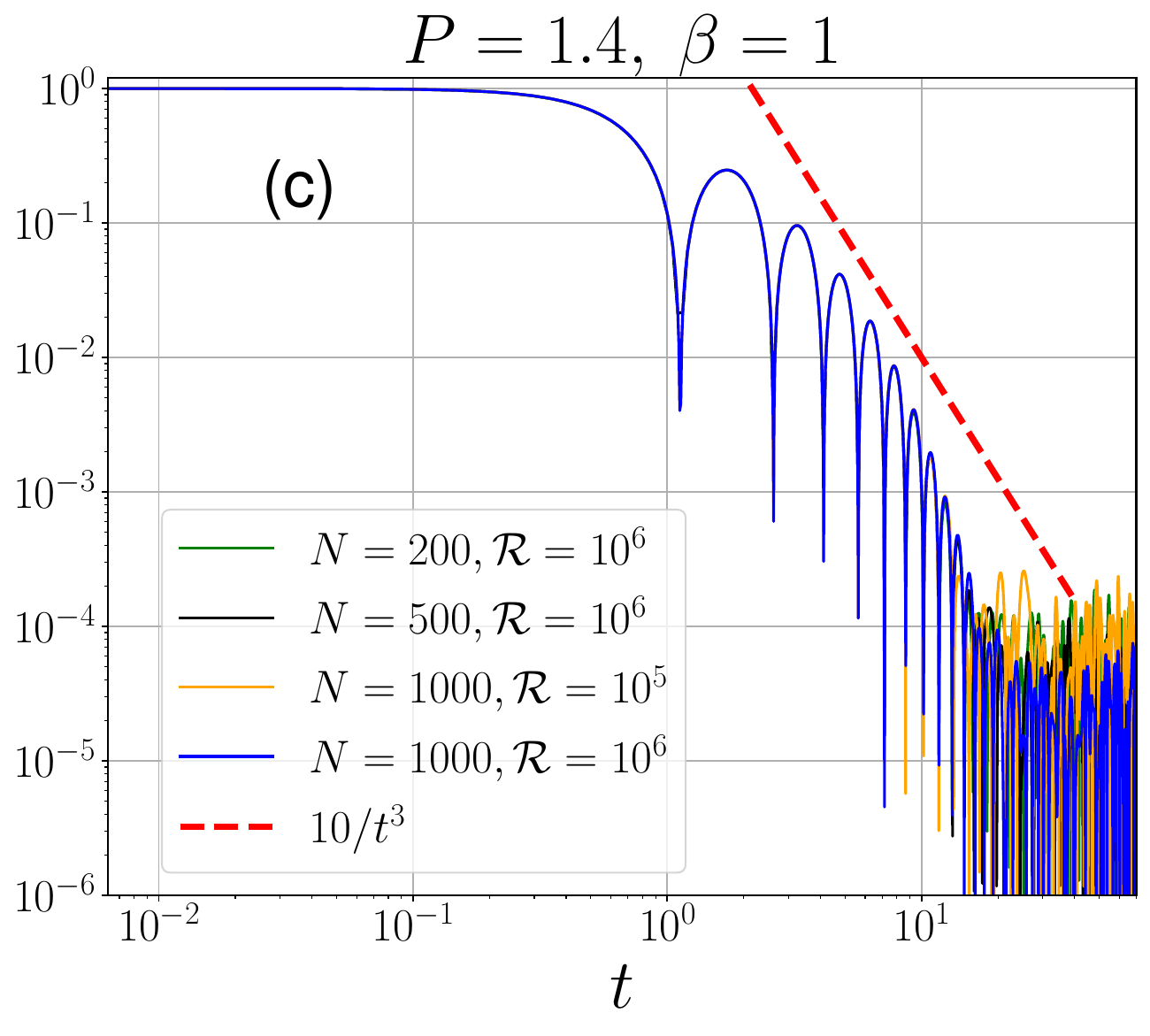}    
    \caption{Plots comparing the velocity autocorrelation function of a tagged particle obtained from simulation (solid lines) with $1/t^3$ scaling (red dashed line) for three set of parameters (a) $P=0.32,~\beta=0.5$, (b) $P=1,~\beta=1$ and (c) $P=1.4,~\beta=1$. The figures demonstrate a slow decay having a power law close to theoretical predictions. The agreement with theory improves upon increasing the system size $N$ and the number of realisations $\mathcal{R}$.}
    \label{fig:p_corr}
\end{figure}

As mention already, for the hard rod fluid the velocity auto-correlation function decays as $t^{-3}$. In fact this function exhibits monotone decay. In Fig. \ref{fig:p_corr} we display our numerical results for the Toda fluid in a log-log plot. One observes rapid oscillations which makes a power law fit somewhat ambiguous. 
The slow decay is well confirmed. The power law seems to be close to $-3$. 

\section{Summary and Outlook}\label{sec8}
For the Toda fluid in thermal equilibrium we numerically confirm that for long times a tagged quasiparticle moves with an average velocity $v^\mathrm{eff}$ and has fluctuations as a Brownian motion. While this behavior has been predicted by GHD, a simulation became possible only through linking the position of the quasiparticle to the respective time-dependent eigenvector of the Lax matrix. As already expected heuristically, in addition we establish that the velocity auto-correlation function of a quasiparticle decays exponentially in time.

We also present similar results for the motion of the eigenvectors. Due to their intrinsic connection with the quasiparticles, they offer an alternative, but equivalent, point of view on the system. The resulting effective velocity and diffusion constant differ quantitatively from those of the quasiparticles. Furthermore, an analytical, GHD-based derivation of the diffusion constant is still missing.

Considering a tagged Toda particle, its velocity vanishes by symmetry. Its fluctuations turn out to be also governed by Brownian motion. As a novel application of GHD, we compute its diffusion constant.
In contrast to the quasiparticle, the velocity auto-correlation function of the tracer particle has a slow decay as $|t|^{-3}$. 

The same scheme can be applied to integrable classical many-body systems possessing a Lax matrix. Prominent examples are the Calogero fluid with
$1/\sinh^2$ repulsive interaction and the Ablowitz-Ladik chain. In the latter case 
required is the underlying particle dynamics as discovered in \cite{brollo24}. Switching
to the continuum limit of the Ablowitz-Ladik chain, the nonlinear Schr\"{o}dinger equation, matters become not so obvious. However very recent progresses is reported for the KdV equation \cite{doyon26,bonnemain26}. Its Lax matrix is the one-dimensional Schr\"odinger equation, whose eigenvectors are still localized in thermal equilibrium. In the case of a pure soliton gas, these eigenvectors appear to play a role similar to the Toda fluid, thereby enabling the tracking soliton trajectories even at high density. 
\section*{Acknowledgements}
AB acknowledge the financial support from the Deutsche Forschungsgemeinschaft (DFG, German Research 
Foundation) – TRR 352 – Project-ID 470903074. AK acknowledges the financial support under project ANRF/ARGM/2025/001207/MTR
from the ANRF, DST, Government of India. AD acknowledges the J.C. Bose Fellowship (JCB/2022/000014) of the Science and Engineering Research Board of the Department of Science and Technology, Government of India. HS thanks the VAJRA faculty scheme (No. VJR/2019/000079) from the Science and Engineering Research Board
(SERB), Department of Science and Technology, Government of India.  SC, IM, AD, and AK would like to acknowledge the support from the DAE, Government of India, under Project No. RTI4001. AD and AK also acknowledge the research support from the International Research Project (IRP) titled ``Classical and quantum dynamics in out of equilibrium systems" by CNRS, France. The authors are thankful to ICTS-TIFR in Bengaluru for the program \textit{``Hydrodynamics, Fluctuations, and Noise in Quantum and Classical Systems"} ( ICTS/hydrodynamics2025/12), during which the early discussions on this work were initiated.  AD, AK and HS  acknowledge discussions held during the program \textit{"Hydrodynamics of low-dimensional interacting systems"} at the Yukawa Institute for Theoretical Physics, Kyoto, during June 2-13, 2025.

\appendix

\section{Lyapunov exponent and Thouless relation}\label{appB}

\textbf{Lyapunov exponent}. To define the Lyapunov exponent we switch to a boundary value problem
on the right half lattice $\mathbb{Z}_+$.
The boundary values are $\psi(0),\psi(1)$, and, since the Lax matrix is tridiagonal,   the eigenvalue equation $L\psi = z \psi$ translates to acting with the transfer matrix 
\begin{equation}\label{9.68}
R(z,j) = \frac{1}{a_j}
\begin{pmatrix}
0 & a_j\\
-a_{j-1} & z - \frac12p_j
\end{pmatrix},
\end{equation}
as
\begin{equation}\label{9.69}
\begin{pmatrix}
  \psi(N)\\
\psi(N+1)
\end{pmatrix} = 
R(z,N)  \cdots R(z,1)
\begin{pmatrix}
  \psi(0)\\
\psi(1)
\end{pmatrix}.
\end{equation}
Asymptotically, the solution increases  exponentially and has a rate determined by
\begin{equation}\label{9.70}
\lim_{N \to \infty} \frac{1}{N} \log\|R(z,N)  \cdots R(z,1)\| = \gamma(z).
\end{equation}
For  thermal equilibrium,  this limit exists almost surely and defines the Lyapunov exponent $\gamma(z)$. On the real axis, $z = \lambda \in \mathbb{R}$, the strict inequality $\gamma(\lambda) >0$ holds. 

The strictly positive Lyapunov exponent is linked 
to Anderson localization. As intuitive picture, an exponential increasing towards the right is matched to an  exponential increasing towards the left, which can be achieved only if $w$ is an eigenvalue with a normalizable eigenvector. Ignoring oscillations, qualitatively an eigenvector of $L$ is then of the form  
$\exp(- \gamma(\lambda)|j-j_\mathrm{max}|)$.

Quite some time ago Thouless \cite{Thouless1972} studied random Jacobi matrices including our $L$ for thermal equilibrium.
He argued for the identity
\be\label{eq:gamma_lambda2}
\gamma(\lambda) = \frac{\nu}{2} + \int d\lambda' \log|\lambda-\lambda'|\rho_\mathrm{DOS}(\lambda').
\ee
This is now known as Thouless relation, which holds for a wide class of random matrices \cite{Thouless1972}.
Remarkably, using  Eq.\eqref{eq_dos_rel}, the Thouless relation is identical to Eq. \eqref{eq_fill1} which originally has been discovered in the context of the thermodynamics for the Lieb-Liniger $\delta$-Bose gas \cite{YangYang1969}.\\\\

\textbf{Numerical study of the Thouless relation}. Since we numerically determined the eigenvectors of $L_N$,
the Thouless relation can be checked as a byproduct. In fact, numerically $|\psi_{\alpha}\rangle$  equals $0$ outside an interval of a few lattice sites. Therefore we use the variance $\sigma^2_\alpha$ of  $|\psi_{\alpha}(j,0)|^2$ as measure of the width of the eigenvector. Comparing with a two-sided pure exponential, one  arrives at $\sigma_\alpha \sim 1/\gamma(\lambda_{\alpha})$. Such approximation is uncontrolled and one can expect only a qualitative agreement. To improve the numerical precision, we actually compute the time-dependent variance $\sigma^2_\alpha(t)$  and then compare with the 
time-averaged variance    

\begin{figure}[t]
    \centering
\includegraphics[width=15.0cm,height=8.0cm]{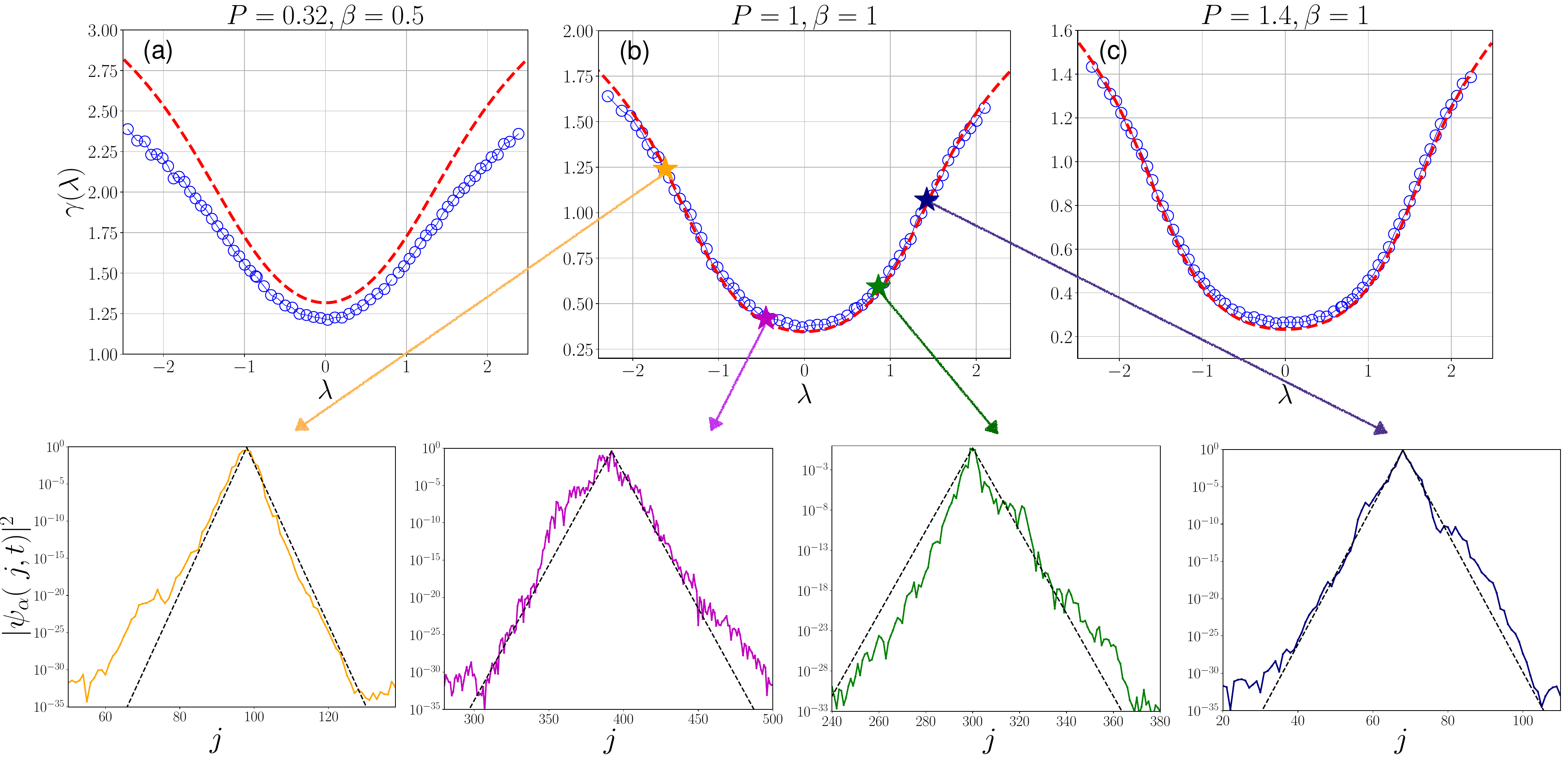}
    \caption{Comparison of the numerically computed Lyapunov exponent $\gamma(\lambda)$, (blue circles) with the analytical prediction from Eq.~\eqref{eq:gamma_lambda2} (red dashed line) for three parameter sets (top panels). Results are shown for a system of size $N=500$, evolved up to $t=2000$ units of time, with $\gamma(\lambda)$ averaged over time and $\mathcal{R}=100$ independent realizations. The lower panels show $\log |\psi_{\alpha}(j,t)|^2$ versus the site index $j$ for selected values of $\lambda$, indicated by colored stars in the top panel, plot (b). The black dashed lines correspond to the theoretical decay $\exp[-2 \gamma(\lambda)|j-j^*|]$, confirming exponential localization. Recall $j^*$ is the argmax of the eigenvector $|\psi_\alpha(j,t)|^2$.}
    \label{fig:Lyapunov}
\end{figure}

\be 
\label{}
\left(\frac{1}{\sigma_\alpha}\right)^{\mathtt r}_{\rm avg}=\frac{1}{t_f}\int_0^{t_f} dt~\frac{1}{\sigma_{\alpha}(t)}.
\ee

As in the previously reported simulations, the eigenvalues $\lambda_\alpha$ is singled out by our binning scheme. Since the eigenvalues vary from one realization to another, repeating the procedure over $\mathcal{R}$ realizations produces a combined dataset of scattered points $\{(\lambda_\alpha^{\mathtt{r}},\left(\frac{1}{\sigma_\alpha}\right)^{\mathtt r}_{\rm avg}\}$ distributed over the spectral axis.  This yields the Lyapunov exponent $\gamma(\lambda)$ as a function of $\lambda$. In Fig.~\ref{fig:Lyapunov}, we compare the numerical results for the Lyapunov exponent $\gamma(\lambda)$ for three different choices of thermodynamic states with the analytical prediction given in Eq.~\eqref{eq:gamma_lambda2}. The numerical solution of Eq.~\eqref{eq:gamma_lambda2} is obtained following the procedure described in Sec.~\ref{sec:TBA-sol}. We observe a good agreement between the numerical data (blue circles) and the theoretical prediction (red dashed line), except in the low-density regime corresponding to $(P=0.32,\beta=0.5)$. To further illustrate the localization properties of the eigenvectors, the insets of Fig.~\ref{fig:Lyapunov}b show the probability density profiles $|\psi_{\alpha}(j,t)|^2$ as a function of the lattice index $j$ for selected values of $\lambda$. The profiles exhibit exponential localization, and their numerical decay is well described by the theoretical form $\exp[-2\gamma(\lambda)|j-j_{\mathrm{max}}|]$. 


\section{Diagonal of the bulk diffusion matrix}\label{appA}

In Sec. \ref{sec3} we showed that the average velocity of a tagged quasiparticle agrees with the effective velocity appearing in the average currents of the hydrodynamic equations. In this appendix we discuss the bulk diffusion matrix $\mathfrak{D}_\mathrm{b}$ \cite{denardis18,denardis19} and establish that its diagonal part matches the self diffusion coefficient $\mathfrak{D}(w)$ computed in Sec. \ref{sec4}. The same observation has already been made in \cite{gopalakrishnan18}.

In the linear response regime, the Onsager matrix is computed via the Green-Kubo formula. Let's recall $J^{[n]}_\mathsf{f}(x,t)$ the current density at $(x,t)$. The Onsager matrix is defined from the microscopic current-current connected correlators as
\be
\mathfrak{L}_{n,m} = \int_0^\infty dt \int dx \left(\langle J^{[n]}_\mathsf{f}(x,t)J^{[m]}_\mathsf{f}(0,0)\rangle^\text{c} - D_{n,m} \right).
\ee
The subtracted term $D_{n,m}=\lim_{t\to\infty}\int dx\langle J^{[n]}_\mathsf{f}(x,t)J^{[m]}_\mathsf{f}(0,0) \rangle^\text{c}$ accounts for the ballistic contribution and it is known as Drude weight and therefore denoted by $D$. Then for us central
quantity is the bulk diffusion matrix, which is denoted by $\mathfrak{D}_\mathrm{b}$ and  
is defined by the relation $\mathfrak{L}=\mathfrak{D}_\mathrm{b}C$, where $C$ is the static covariance matrix $C_{n,m} = \int dx\langle Q^{[n]}_\mathsf{f}(x)Q^{[m]}_\mathsf{f}(0)\rangle^\text{c}_{P,V}$, see Eq. \eqref{eq_corr}. 

Although generally hard to compute, the current-current correlator can be expanded in form factors, and integrability allows the computations of those. This was done in Ref.\cite{denardis18,denardis19} for the Lieb-Liniger model, but  is believed to be valid in any integrable model.  To state the result, let us define the kernel
\be
K(w, w') = \rho_{\mathsf{p}}(w)\rho_{\mathsf{p}}(w')f(w') |v^{\text{eff}}(w) - v^{\text{eff}}(w')| |\varphi\dr(w, w')|^2,
\ee
where $\varphi\dr=\left[1 - \varphi\vartheta \right]^{-1}\varphi$ is the dressed scattering kernel. Then the Onsager matrix takes the form 
\be\label{eq_onsanger}
\mathfrak{L}_{n,m} = \frac{1}{2} \int~\mathrm{d}w \mathrm{d}w' K(w, w') \left( \frac{(q^{[n]})\dr(w')}{\rho_{\mathrm{s}}(w')} - \frac{(q^{[n]})\dr(w)}{\rho_{\mathrm{s}}(w)} \right) \left( \frac{(q^{[m]})\dr(w')}{\rho_{\mathrm{s}}(w')} - \frac{(q^{[m]})\dr(w)}{\rho_{\mathrm{s}}(w)} \right),
\ee
with $q^{[n]}(w)$ the rapidity resolved charge densities.
The bulk diffusion matrix is extracted from $(\mathfrak{D}_\mathfrak{b})_{n,m}=(\mathfrak{L}C^{-1})_{n,m}$. In this way, we obtain the diffusion matrix in the charge basis. However, to compare it with the self diffusion in Eq. \eqref{eq:mfrk(D)}, we need to express it in the rapidity basis. Let's rewrite the Onsager matrix as an operator $\mathfrak{L}_{n,m} = \langle q^{[n]}, (\mathfrak{D}_\mathrm{b}C)q^{[m]}\rangle$, with
\be\label{eq_onsop}
\begin{split}
&\mathfrak{D}_\mathrm{b}C = [1-\vartheta T]^{-1}\mathfrak{\tilde L}[1-T\vartheta]^{-1}, \\ 
&\mathfrak{\tilde L}(w,w') = \rho_\mathsf{s}^{-1}(w)\rho_\mathsf{s}^{-1}(w') \left( \delta(w-w')\int~dw''K(w,w'') - K(w,w')\right),
\end{split}
\ee
where $[1-T\vartheta]^{-1}$ is the dressing operator in \eqref{eq_dress} and $\mathfrak{\tilde L}(w,w')$ is the integral kernel of the operator $\mathfrak{\tilde L}$. Similarly, from Eq. \eqref{eq_corr} we have 
\be
C = [1-\vartheta T]^{-1}\rho_\mathsf{p}[1- T\vartheta]^{-1}.
\ee
After simple algebraic manipulations, we obtain
\be\label{eq_diffker}
\mathfrak{D}_\mathrm{b} = [1-\vartheta T]^{-1}\rho_\mathsf{s}\mathfrak{\tilde D}_\mathrm{b}\rho_\mathsf{s}^{-1}[1-\vartheta T] = R^{-1}\mathfrak{\tilde D}_\mathrm{b}R,
\ee
with $R=\rho_\mathsf{s}^{-1}[1-\vartheta T]$. As we showed in Sec.\ref{sec4}, this matrix is the Jacobian for the transformation from the quasiparticles density to the filling function $\delta\vartheta(w)=R\delta\rho_\mathsf{p}(w)$. This matrix diagonalizes the flux Jacobian, so we can interoperate it as the change of basis towards the hydrodynamics normal modes (see \cite{doyon20}, Section 3.5, for full discussion).
As a consequence, we get the rapidity resolved bulk diffusion from the integral kernel $\mathfrak{\tilde D}_\mathrm{b}(w,w')$. This is explicitly given by
\be
\mathfrak{\tilde D}_\mathrm{b}(w,w') = \frac{1}{\rho_\mathsf{s}(w)^2}\left( \delta(w-w')\int~dw''K(w,w'') - K(w,w')\right)\frac{1}{\rho_\mathsf{p}(w')}.
\ee
Remarkably, the diagonal part gives the TBA prediction for the self-diffusion 
defined in Eq.~\eqref{eq:mfrk(D)}. This is explicitly
\be\label{eq_tbabd}
\mathfrak{\tilde D}_\mathrm{b}(w) =  \int~dw' \rho_\mathsf{p}(w')\left( \frac{\varphi\dr(w,w')}{\rho_\mathsf{s}(w)} \right)^2|v^\text{eff}(w)-v^\text{eff}(w')|.
\ee

\bibliography{refs.bib}

\nolinenumbers

\end{document}